\documentclass[12pt]{iopart}
\usepackage[utf8]{inputenc}
\usepackage{amssymb,amsfonts,setspace,graphicx,bm,float,amssymb,siunitx}

\newcommand{\bra}[1]{\langle #1|}
\newcommand{\ket}[1]{|#1\rangle}

\newcommand{\ketbra}[2]{| #1 \rangle \langle #2 |}
\newcommand{\expect}[1]{\langle #1 \rangle}

\expandafter\let\csname equation*\endcsname\relax

\expandafter\let\csname endequation*\endcsname\relax
\usepackage{amsmath}

\begin{document}
\title{Designing Kerr interactions using multiple superconducting qubit types in a single circuit}

\date{\today}

\author{Matthew Elliott} \address{Advanced Technology Institute and Department of Physics, University of Surrey, Guildford, GU2 7XH, United Kingdom}
\author{Jaewoo Joo}
\address{School of Computational Sciences and School of Electronics, Korea Institute for Advanced Study, Seoul 02433, Republic of Korea}
\address{Clarendon Laboratory, University of Oxford, Parks Road, Oxford, OX1 3PU, United Kingdom}
\author{Eran Ginossar} \address{Advanced Technology Institute and Department of Physics, University of Surrey, Guildford, GU2 7XH, United Kingdom}

\begin{abstract}
The engineering of Kerr interactions has great potential for quantum information processing applications in multipartite quantum systems and for investigation of many-body physics in a complex cavity-qubit network. We study how coupling multiple different types of superconducting qubits to the same cavity modes can be used to modify the self- and cross-Kerr effects acting on the cavities and demonstrate that this type of architecture could be of significant benefit for quantum technologies.

Using both analytical perturbation theory results and numerical simulations, we first show that coupling two superconducting qubits with opposite anharmonicities to a single cavity enables the effective self-Kerr interaction to be diminished, while retaining the number splitting effect that enables control and measurement of the cavity field. We demonstrate that this reduction of the self-Kerr effect can maintain the fidelity of coherent states and generalised Schr\"{o}dinger cat states for much longer than typical coherence times in realistic devices. Next, we find that the cross-Kerr interaction between two cavities can be modified by coupling them both to the same pair of qubit devices. When one of the qubits is tunable in frequency, the strength of entangling interactions between the cavities can be varied on demand, forming the basis for logic operations on the two modes. Finally, we discuss the feasibility of producing an array of cavities and qubits where intermediary and on-site qubits can tune the strength of self- and cross-Kerr interactions across the whole system. This architecture could provide a way to engineer interesting many-body Hamiltonians and a useful platform for quantum simulation in circuit quantum electrodynamics.

\end{abstract}

\maketitle

\section{Introduction}
Experimental progress in coherent superconducting circuits has resulted in a wide variety of qubit designs, from the prototypical flux \cite{Friedman2000, VanderWal2000, Yan2016, Steffen2010}, phase \cite{Martinis2002,Ansmann2009, Strauch2003, Chen2012} and charge \cite{Nakamura1999,Bouchiat1998, Pashkin2009} qubits to more modern designs such as transmon \cite{Koch2007} and fluxonium \cite{Manucharyan2009} circuits. The mathematical description and quantum dynamics of qubits when coupled transversely to resonators is provided by circuit quantum electrodynamics (cQED). While these devices are all designed to approximate a two-level system in a superconducting circuit, they all possess many additional levels which can affect their interactions with other elements. In particular, when a qubit is coupled to a linear resonator these extra levels can determine the type of nonlinearity that is induced in that resonator. When a superconducting qubit is far detuned from a resonator mode but is nonetheless strongly coupled to it, the second order effect of this induced nonlinearity is known as the Kerr effect. In superconducting devices the Kerr interaction can be significant on the single-photon level \cite{Kirchmair2013}, a regime which is difficult to reach in optical atom-cavity systems. While the nonlinearity induced by a qubit, via state dependent shifts, enables information to be encoded and controlled in coherent quantum states of a cavity mode, the Kerr effect also distorts these states and rapidly reduces their fidelity over time \cite{Leghtas2013}. The strength of the Kerr effect is therefore an important property of the resonator mode. Here we study how the induced Kerr effects can be engineered by combining different types of superconducting qubits in a single circuit. The existence of mature circuit designs at different experimental groups suggests that implementing this type of design, which two different superconducting qubits on a single substrate, might now be feasible.

While longitudinal coupling between devices is increasingly being studied in superconducting circuits \cite{Richer2016,Kerman2013,Wang2016,Vacanti2012}, we will consider only consider only the transverse coupling described by cQED. In this configuration, coupled qubits can give rise to Kerr interactions both on single cavities and between pairs of cavities. The self-Kerr is a nonlinear shifting of a resonator frequency as a function of the number of photons in the mode. A simple quantum system where this can been seen is the quantum Duffing oscillator, with its term proportional to $(a^\dag a)^2$ in the Hamiltonian, where $a$ is the photon annihilation operator for a resonator mode. In the classical limit, this becomes the quadratic dependence of the refractive index on the electric field strength, sometimes known as self-phase modulation \cite{Drummond1980a,Drummond1980}. This effect manifests itself at the second order in a series expansion of the Jaynes-Cummings interaction in the dispersive limit \cite{Boissonneault2009}. With two resonators, this idea can be extended to the cross-Kerr effect, known as cross-phase modulation in optics, which is indicated by a term proportional to $a^\dag a b^\dag b$. Cross-Kerr can be realised by coupling two cavities of different frequency via a single qubit device. Both of these effect are also of interest in a variety of analogous quantum systems, such as cavity optomechanics \cite{Khan2015, Xiong2016}, where large nonlinearities can feasibly be produced.

In optical systems Kerr effects are typically very weak due to the weak coupling that can be realised between natural atoms and optical modes. By contrast, in cQED the strong couplings in the dispersive regime mean that very strong self- and cross-Kerr effects can be produced, and both nonlinearities have been demonstrated on the level of single photons in recent experiments \cite{Schuster2007, Kirchmair2013, Holland2015}. It is known that this large interaction strength \cite{Rebic2009} is required to implement logic qubit gates using the cross-Kerr interaction \cite{Chuang1995, Milburn1989}, while strong self-Kerr can be used to generate Schr\"{o}dinger cat states in cavity \cite{Glancy2008}. Recent developments in three-dimensional microwave cavities have produced coherence times of nearly 1ms \cite{Reagor2016}, leading to great interest in using the cavity field to store continuous-variable quantum information in Schr\"{o}dinger cat states \cite{Mirrahimi2014}. Specific protocols to prepare \cite{Vlastakis2013}, encode quantum information in \cite{Leghtas2013}, stabilise \cite{Leghtas2015}, entangle \cite{Wang2016} and error correct \cite{Ofek2016} such cavity states have been demonstrated experimentally. Recent work also discusses how to transmit generalised cat states out of a cavity into a transmission line \cite{Pfaff2016}. In many cases, for example if trying to build a long-lived quantum memory, it is desirable to have very weak nonlinearites so that stored states can be preserved with high fidelity. A method to design different combinations of Kerr nonlinearities, and even vary them dynamically could therefore be extremely useful.

In this paper we demonstrate that, by combining different types of superconducting qubits with different anharmonicities in a single circuit, the strength of both the self- and cross-Kerr interactions can be designed. In addition, the use of tunable qubits allows these interactions to be controlled dynamically to a much larger extent than is possible in circuits based on one type of qubit. We focus on dynamics of quantum states in the single photon regime and well within the coherent timescales where quantum information processing is designed to operate. In Section \ref{sec2} we start by considering modifying self-Kerr by coupling two different superconducting qubits to a single mode. As coherence times improve, self-Kerr becomes more significant in relation to the cavity decay rate and introduces larger phase distortions to these cavity states, reducing the fidelity of any stored quantum state. In Section \ref{sec2-1} a time-independent perturbation theory is used to show that the self-Kerr interaction can be passively eliminated from the cavity field while still retaining the number splitting required for controlling and measuring the cavity. In Section \ref{sec2-2} we show that this result persists when a more realistic model of transmon qubits, the quantum Duffing oscillator with several levels, is considered. Using exact numerical simulation of the Schr\"odinger equation, we show that the perturbation theory method well describes the behaviour of the full quantum system. In Section \ref{sec2-3} we demonstrate numerically that this self-Kerr cancellation is also possible in the case where the qubit, in this case fluxonium \cite{Manucharyan2009} cannot be modelled as a Duffing oscillator. These results suggest that such a setup could be used to increase the fidelity of quantum memories. A similar cancellation idea has recently been demonstrated using a single multi-level system in cQED \cite{Juliusson2016}.

In Section \ref{sec3}, we extend the principle to modifying the cross-Kerr interaction between two cavities. The cross-Kerr effect is of particular interest in quantum optics because of its ability to entangle two optical degrees of freedom and therefore has the potential to be used for quantum information processing. The interaction can be used to perform entangling gates between qubits encoded in superpositions of coherent states or Fock states depending on the computational basis that is being used. For travelling photonic states at optical frequencies it is difficult to achieve sufficient strong nonlinearities to perform the entanglement of any reasonable timescale. However this can be significantly improved in trapped photonic states in microwave circuit QED. If two modes, both starting in the state $a_i\ket{0}+b_i\ket{1}$, interact by the cross-Kerr interaction for a time $t$, the resulting state is
\begin{equation}
      e^{ig\, a^\dag a \, b^\dag b \, t} (a_1\ket{0}+b_1\ket{1})(a_2\ket{0}+b_2\ket{1}) = a_1a_2\ket{0}\ket{0}+a_2b_1\ket{1}\ket{0}+a_1b_2\ket{0}\ket{1}+ e^{igt}b_2b_1\ket{1}\ket{1}.
\end{equation}
The interaction only acts on the final term of the superposition as it has no effect if there is a vacuum state in either mode, and waiting a time $t=\pi/g$ will therefore change the sign of this term. This realises a controlled-phase gate ($e^{i\pi\, a^\dag a \, b^\dag b }$) to obtain a maximally entangled state between two modes. A similar scheme has recently been theoretically proposed for entangling propagating photons \cite{Brod2016}. We consider two cavities which are coupled together by two intermediary qubits with opposite anharmonicities. In this configuration we can realise two different regimes. In one regime the cross-Kerr interaction is entirely cancelled by the presence of a second superconducting qubit providing excellent mutual isolation of the cavities (Section \ref{sec3-1}). In the other regime one of the superconducting qubits is moderately detuned and the cross-Kerr produces a maximally entangled state between the cavities (Section \ref{sec3-2}). Switching between these two configurations through the use of a tunable qubit allows an entangling gate to be performed and then the interaction switched off to maintain this entangled state, forming the basis of a quantum logic gate.

Finally, in Section \ref{sec4} we consider how this Kerr engineering could scale up to many-body systems. Circuit QED has opened avenues in many-body physics by providing a potential way to engineer an array of cavities and using superconducting qubits to modify the on-site and coupling parameters. SQUID loops have also been studied as a way to induce correlations between quantum resonators through the use of on-site and intermediate devices \cite{Stassi2015}. This kind of setup is of interest for studying phenomena such as quantum phase transitions \cite{Georgescu2014, Jin2013}, quench dynamics \cite{Creatore2014} and for implementation of a quantum simulator \cite{Houck2012} in a driven-dissipative system. It is theoretically possible to realise an analogue of the Bose-Hubbard model \cite{Gersch1963,Freericks1994} in circuit QED using coupling to qubits to provide the on-site nonlinearity, known as the Jaynes-Cummings lattice \cite{Leib2010,Zhu2013, Nissen2012}. In other work, the phase diagram of a line of cavities possessing only cross-Kerr interactions has also been solved theoretically \cite{Jin2014}, and a very recent experiment reports a quantum phase transition in a line of 72 superconducting cavities \cite{Fitzpatrick2016}. We here consider a line of cavities, each with an on-site qubit and coupled by intermediary qubits along the line and show that, in principle, the use of qubits with different anharmonicities in this setup could allow the realisation of novel many-body Hamiltonians. The tunable qubits would also enable the study of quantum quenches and other interesting phenomena \cite{Calabrese2007}, where parameters of the system Hamiltonian are changed suddenly, revealing information about many-body systems.

\section{Modifying cavity self-Kerr for quantum memory}
\label{sec2}
The self-Kerr induced by a qubit on a linear resonator can be derived from diagonalising the Jaynes-Cummings (JC) Hamiltonian
\begin{equation}
    H^{JC} = \omega_c a^\dag a +\frac{\omega_q}{2} \sigma_z + g\left(a\sigma_+ +a^\dag \sigma_- \right),
\end{equation}
where $a$ is the cavity annihilation operator, $\sigma_\pm$ are the qubit raising and lowering operators, $\omega_c$ is the cavity frequency, $\omega_q$ is the qubit frequency and $g$ is the cavity-qubit coupling. The JC model assumes that the rotating wave approximation (RWA) \cite{Zueco2009} holds and that interactions that create excitations simultaneously in the cavity and transmon are therefore very unlikely. This means that the model is only valid when $|\omega_c - \omega_q|\ll |\omega_c + \omega_q|$. Outside of this parameter range, the Rabi model is required \cite{Rabi1937}, giving rise to interesting new physics \cite{Irish2005}. We discuss the breakdown of the RWA in the context of our results in Appendix \ref{app:RWA}. The eigenvalues of the JC model can be solved for exactly by diagonalising the Hamiltonian in the $\ket{n,\uparrow},\ket{n+1,\downarrow}$ basis, with adjacent blocks uncoupled because the total number of excitations conserved by the interaction. The Hamiltonian can be written in $2 \times 2$ matrix blocks
\begin{equation}
    H_n^{JC}=\left(\begin{matrix}
  n\omega_c + \frac{\omega_q }2 & g\sqrt{n+1}   \\
  g\sqrt{n+1} & (n+1)\omega_c -\frac{\omega_q}{2}
 \end{matrix}\right),
\end{equation}
where $n+1$ is the total number of excitations in the system ($0\le n\le N$). Separately, we also have the ground state $\ket{0,\downarrow}$ and the state $\ket{N,\uparrow}$ which has the maximum number of excitations permitted in the chosen basis. As the blocks are not coupled, we can diagonalise them separately, giving the eigenvalues
\begin{equation}
\label{eqn:eigenvalues}
    E_n^{JC}=\left(n+\frac12\right)\omega_c \pm \frac{1}{2} \sqrt{4g^2(n + 1) + \Delta^2 },
\end{equation}
where $\Delta = \omega_c -\omega_q$ is the detuning between the cavity and qubit. The eigenstates of the system are
\begin{align}
    \ket{n,+} = \cos\left(\frac{\theta_n}2\right) \ket{n+1,\downarrow} + \sin\left({\frac{\theta_n}2}\right)\ket{n,\uparrow}, \label{dressed1}
    \\
    \ket{n,-} = -\sin\left(\frac{\theta_n}2\right) \ket{n+1,\downarrow} + \cos\left({\frac{\theta_n}2}\right)\ket{n,\uparrow},\label{dressed2}
\end{align}
with $\theta_n=\arctan(2g \sqrt{n+1}/\Delta)$. In the dispersive regime $\Delta \gg g$, we can expand the square root term of Eqn. \ref{eqn:eigenvalues} in the small parameter $g/\Delta$ and assume that the qubit always remains in its ground state. This gives use the approximate eigenvalues

\begin{equation}
     E_n^{JC} \approx  \left(\omega_c-\frac{g^2}{\Delta}+\frac{2g^4}{\Delta^3} \right)n + \frac{g^4}{\Delta^3} n^2.
\end{equation}
The nonlinear $n^2$ term describes the cavity self-Kerr, the sign of which is completely determined by the sign of the detuning. This dispersive energy spectrum can also found by approximately diagonalising the JC Hamiltonian by application of an appropriate unitary transformation \cite{Boissonneault2009,Carbonaro1979} or by calculating a perturbation expansion up the fourth order in the interaction \cite{Meystre}.

\subsection{Eliminating cavity self-Kerr with a pair of two-level qubits}
\label{sec2-1}
A single qubit induces a self-Kerr interaction proportional to $g^4/\Delta^3$ on the cavity, which can be modified by changing either the coupling or the detuning between the qubit and the cavity. Detuning the qubit very far from the cavity, however, turns off the number splitting required for control and read out of the cavity field. We therefore add a second qubit and investigate whether this can be used instead to modify the nonlinearity of the cavity. The Hamiltonian for the one-cavity and two-qubit (1C2Q) system is given by
\begin{equation}
      H^{1C2Q} = \omega_c a^\dag a +\frac{\omega_{q1}}{2} \sigma_{z1} +\frac{\omega_{q2}}{2} \sigma_{z2}+ g_1\left(a\sigma_{+1} +a^\dag \sigma_{-1} \right)+ g_2\left(a\sigma_{+2} +a^\dag \sigma_{-2} \right),
\end{equation}
where $g_1$ and $g_2$ are the couplings to the two qubits, with their respective detunings $\Delta_1=\omega_c-\omega_{q1}$ and $\Delta_2=\omega_c-\omega_{q2}$. If $\Delta_1=\Delta_2$ this is known as the two-atom Dicke model \cite{Garraway2011}. We can also diagonalise this system block-wise in $4 \times 4$ blocks given by

\begin{equation}
    H_n^{1C2Q}=\left(\begin{matrix}
  (n+1)\omega_c + \frac{\Delta_+}2 & g_1\sqrt{n+2} &  g_2\sqrt{n+2} & 0 \\
  g_1\sqrt{n+2} & (n+1)\omega_c- \frac{\Delta_- }2 & 0 & g_2\sqrt{n+1}\\
  g_2\sqrt{n+2} & 0 & (n+1)\omega_c+ \frac{\Delta_- }2 & g_1\sqrt{n+1} \\
  0 & g_2\sqrt{n+1} & g_1 \sqrt{n+1} & (n+1) \omega_c -  \frac{\Delta_+}2
 \end{matrix}\right),
\end{equation}
for $\Delta_+ = {\Delta_1}+{\Delta_2}$ and $\Delta_- = {\Delta_1}-{\Delta_2}$.
The analytical form of the eigenvalues of this block is very complicated in general, but if we consider the case $g_1=g_2=g$, and $\Delta_1=-\Delta_2=\Delta$, then the eigenvalues are
\begin{equation}
\begin{matrix}
    (n+1)\omega_c& ~~{\ket{n, \downarrow \downarrow}~\rm or~\ket{n, \uparrow \uparrow}}~,\\
(n+1)\omega_c \pm \sqrt{4g^2 (n+{\frac{3}{2}})+\Delta^2} & ~~{\ket{n, \uparrow \downarrow}~\rm or~\ket{n, \downarrow \uparrow}}~.
\end{matrix}
\end{equation}
As for the single-qubit case in Equations (\ref{dressed1}) and (\ref{dressed2}), the eigenvectors in the dispersive regime are close to being the bare eigenstates. The first eigenvalue applies to the states where the qubits are both in the same state. This means that, in the dispersive regime, if both qubits remain in the ground state the cavity self-Kerr is completely eliminated. The cavity state feels no nonlinearity, but the number splitting of the qubit transition frequency remains, which allows the qubit to still be used to control and measure the cavity. This result is applicable to the flux qubits used in a recent experiment, which have strong coupling and very large anharmonicities and can therefore be considered as close to a true two-level system \cite{Stern2014}.

\begin{figure}
    \centering
    \includegraphics[width=0.5\columnwidth]{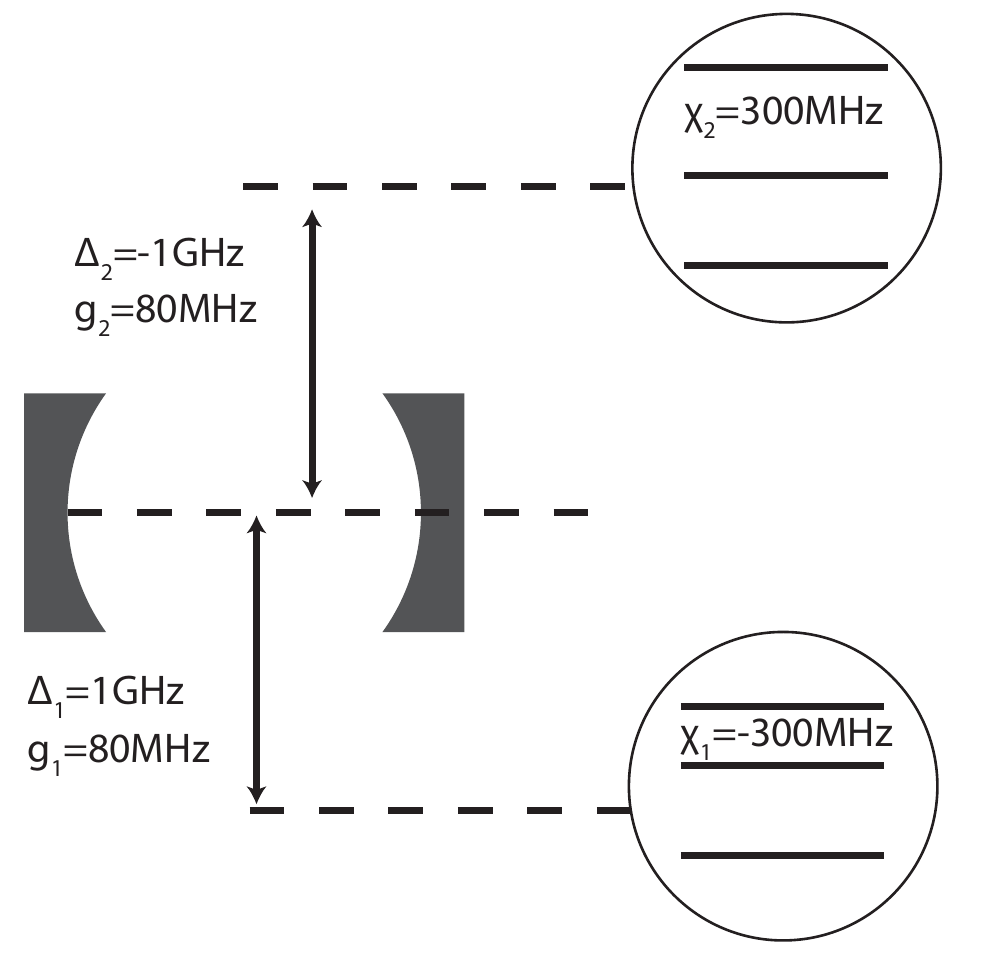}
    \caption{Schematic of one cavity and two qubit setup that achieves cancellation of the cavity self-Kerr effect. The two Duffing-like devices have both their anharmonicities and detunings from the cavity of equal magnitude but with opposite signs. The cavity field can still be controlled via the qubit using number splitting of the qubit frequency}
    \label{fig:selfkerr}
\end{figure}

\subsection{Eliminating self-Kerr with two Duffing-like qubits}
\label{sec2-2}
In practice, many superconducting qubits are not ideal two-level systems, but are instead relatively weakly anharmonic oscillators. The transmon \cite{Koch2007}, for example, can be modelled as a quantum Duffing oscillator \cite{Boissonneault2009} with a constant anharmonicity when only the lowest few levels are relevant. With many qubit levels the exact diagonalisation used above becomes increasingly difficult, but we can still use perturbation theory to derive expressions for the eigenenergies in the dispersive limit. The Hamiltonian for a cavity coupled to a single transmon is
\begin{equation}
      H^{1C1T} = \omega_c a^\dag a + \omega_q b^\dag b + \chi b^\dag b^\dag b b + g(ab^\dag+a^\dag b),
\end{equation}
where the transmon field has annihilation operator $b$ and nonlinearity $\chi$. To calculate perturbation theory results that include the self-Kerr, we need to calculate terms up to the fourth order in the interaction Hamiltonian. The presence of the third transmon level adds a correction to the energies associated with two excitations moving from the cavity into the qubit (full details given in \ref{app:selfcross}), giving the modified cavity eigenenergies
\begin{equation}
    E_{n,g}^{1C1T}= \left[\omega_c + \frac{g^2}{\Delta} -\frac{2g^4}{\Delta^2(2\Delta+\chi)}\right]n + \frac{\chi g^4}{\Delta^3(2\Delta + \chi)}n^2,
\end{equation}
to fourth order in $g/\Delta$. We can see that, with a finitely anharmonic qubit, the sign of the nonlinearity induced on the cavity is not fully determined by $\Delta$, but also depends on the sign of $\chi$ \cite{Murch2012}. Two qubit levels are therefore only sufficient to predict even the sign of the interaction when $|\chi|\gg|\Delta|$, which is not generally satisfied in the dispersive regime of cQED.  While $\chi$ is always negative in transmon devices, flux qubits can behave like a Duffing oscillator in the absence  of external flux and produce energy levels with a positive anharmonicity of a few hundred megahertz \cite{Yan2016}.

Next, we consider coupling a second Duffing-like qubit to the same cavity. The total Hamiltonian is equivalent to
\begin{equation}
            H^{1C2D} = \omega_c a^\dag a + \sum_{i=1}^2\left[ \omega_i b_i^\dag b_i + \chi_i b_i^\dag b_i^\dag b_i b_i + g_i(ab_i^\dag+a^\dag b_i)\right],
\end{equation}
and the eigenenergies are
\begin{equation}
      \begin{split}
          E_{n,g}^{1C2D}=& \left( \omega_c -\frac{g_1^2g_2^2(\Delta_1+\Delta_2)}{\Delta_1^2\Delta_2^2} + \sum_{i=1}^2 \left( \frac{g_i^2}{\Delta_i} -\frac{2g_i^4}{\Delta_i^2(2\Delta_i+\chi_i)} \right) \right)n + \sum_{i=1}^2  \frac{\chi_i \, g_i^4}{\Delta_i^3(2\Delta_i + \chi_i)} n^2.
      \end{split}
\end{equation}

To cancel the cavity self-Kerr with the additional qubit levels included, we now need to choose $\Delta_1=-\Delta_2$ and $\chi_1=-\chi_2$. We therefore need to use two different types of qubit in a single circuit with opposite nonlinearities, for example transmons and the flux qubits discussed above. We must also ensure that when designing the circuit we choose the parameters such that $g_i\ll|\omega_c - \omega_i|\ll|\omega_c + \omega_i|$ so that we remain within the dispersive regime and RWA.

\begin{figure}[h!!]
    \centering
    \includegraphics[width=0.5\columnwidth,clip=true,trim=1cm 0 1.7cm 1.2cm]{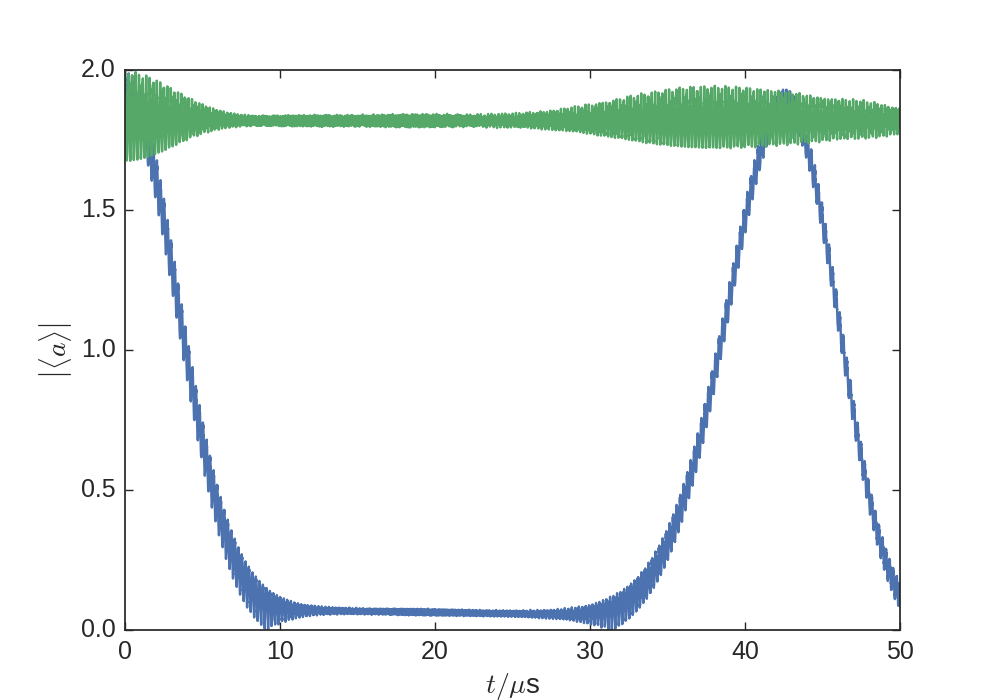}
    \caption{Plot of coherent amplitude $|\expect{a}|$ as a function of time for an initial coherent state with $\alpha=2.0$ showing complete cancellation of the self-Kerr in the 1C2D system. The evolution in the presence of a single transmon is shown in blue and dynamics with both a transmon and a flux qubit modelled by a positively anharmonic Duffing oscillator are shown in green. System parameters are $\Delta_1=-\SI{1}{\giga\hertz}$, $\Delta_2=\SI{1}{\giga\hertz}$, $\chi_1=\SI{-300}{\mega\hertz}$, $\chi_2=\SI{300}{\mega\hertz}$, and $g_1=g_2=\SI{100}{\mega\hertz}$. With a single transmon, the amplitude oscillates due to the cavity self-Kerr. This is completely eliminated by the addition of the second device and the amplitude remains constant, apparently indicating that all orders of the anharmonicity are cancelled. The small reduction in amplitude and the faster oscillations are due to the fact that the eigenstates of the total system are not pure cavity eigenstates.}
    \label{fig:transtrans}
\end{figure}

We can numerically solve the Schr\"odinger equation to see whether this result holds when all orders of $H^{1C2T}$ are included. The action of the Kerr effect on a coherent state is to distort the state, transiently producing various coherent state superpositions before completing a full revival of the initial state \cite{Kirchmair2013}. This gives rise to a periodic collapse and revival of the cavity amplitude $|\expect{a}|$ when starting with a coherent initial condition. In Figure \ref{fig:transtrans} we show that our two-qubit setup appears to cancel not just the cavity self-Kerr, but all orders of the cavity nonlinearity when the detunings and anharmonicities are equal and opposite. This occurs in this idealised case because the energy spectrum is completely symmetrical in the frame of the cavity. In practise, the extent to which the interaction can be reduced may be limited by small non-RWA effects, as discussed in \ref{app:RWA}. We neglect decoherence processes as self-Kerr only has significant effects when it is many times greater than the dissipation and the effects we are looking at therefore occur on much shorter timescales than the dissipation. We see that a single transmon causes the periodic revivals expected, while after the addition of the second qubit an initial coherent state undergoes almost trivial evolution. Some high frequency oscillations appear, caused by the fact that the bare cavity eigenstates are not eigenvalues of the full coupled system. This demonstrates that different types of qubit used together in the same circuit can be used to achieve modifications of the Kerr effect that are not possible using transmons only, in this case with potential to improve the quality of quantum memory.

\subsection{Improved quantum memory using fluxonium qubits}
\label{sec2-3}
To demonstrate that this principle applies to other qubit devices, even those that cannot be modelled as a Duffing oscillator, we also show simulations using a fluxonium qubit \cite{Manucharyan2009} to cancel the cavity self-Kerr, as was proposed in an earlier paper \cite{Joo2015}. The transmon device provides a negative nonlinearity and can be used to control the cavity, while we use the fluxonium to provide an opposite, positive nonlinearity. While our perturbation theory results do not apply fully to this system, as fluxonium has nonzero matrix element for all qubit transitions, rather than just adjacent levels, we see that significant cancellation can still be achieved.
\begin{figure}[h!!]
    \centering
    \includegraphics[width=.5\columnwidth,clip=true,trim=1cm 0 1.7cm 1.2cm]{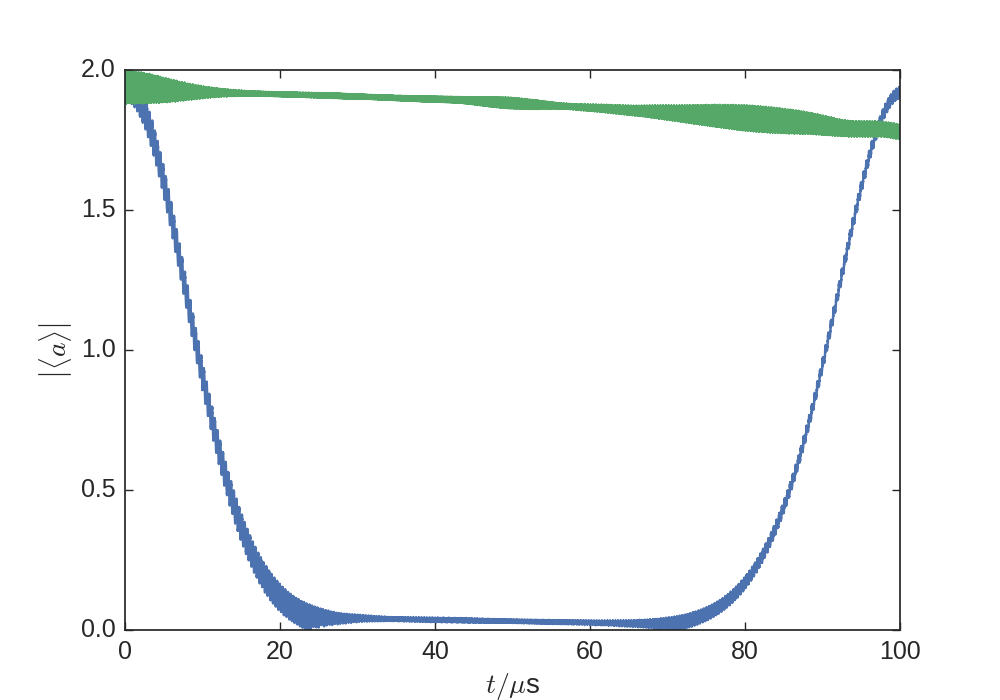}
    \includegraphics[width=.9\columnwidth]{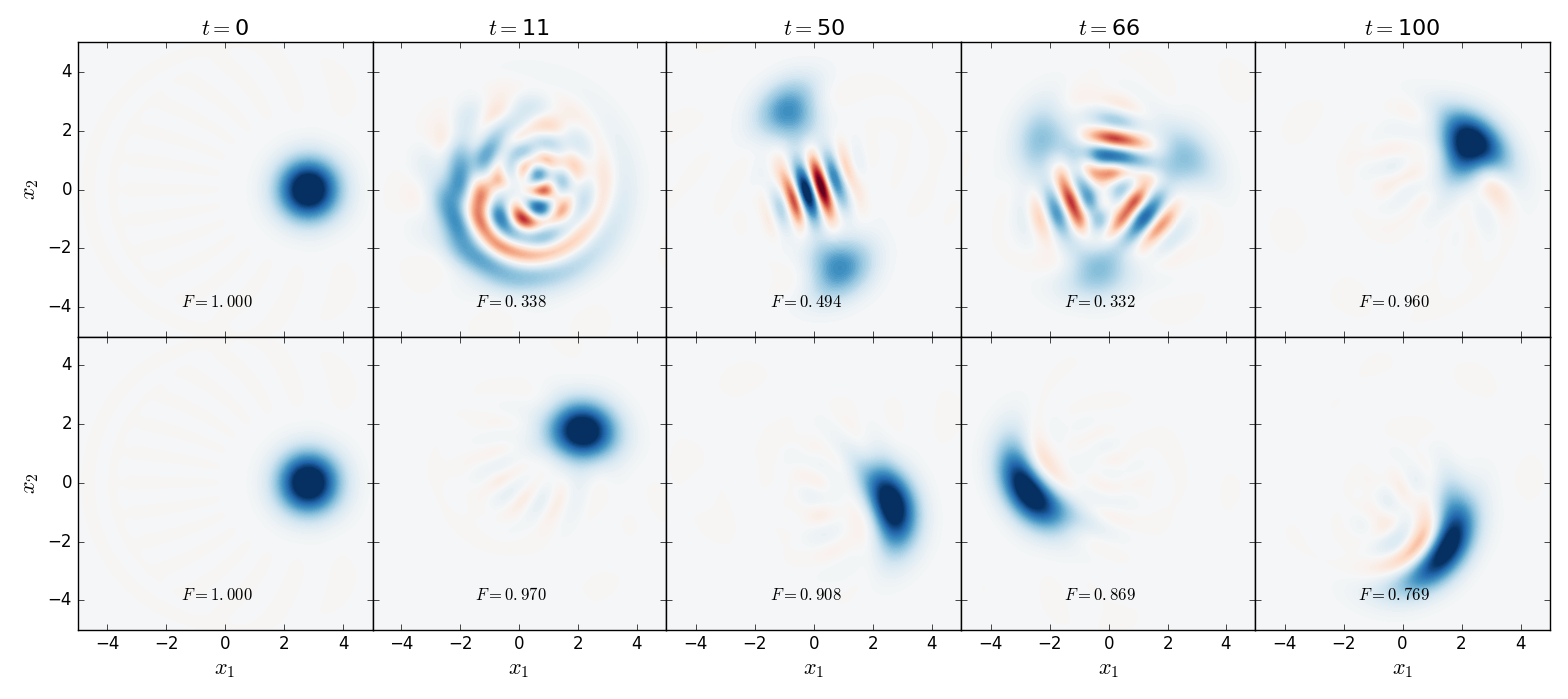}
    \caption{(a) Plot of coherent amplitude $|\expect{a}|$ as a function of time for an initial coherent state with $\alpha=2.0$ showing significant cancellation of self-Kerr by a fluxonium. Evolution in the presence of a transmon only in blue and both a transmon and a fluxonium in green. With only a transmon, the amplitude oscillates due to the cavity self-Kerr, which is greatly reduced by the addition of the fluxonium, which adds a positive anharmonicity to the cavity. $\omega_c=\SI{9.2}{\giga\hertz}$, $\omega_t=\SI{8.2}{\giga\hertz}$, $ \chi=\SI{-300}{\mega \hertz}$, $g_t=\SI{80}{\mega\hertz}$, $\omega_{f,j}=0, 5.505, 10.904, 13.213\si{\giga\hertz}$ and $g_f=\SI{30}{\mega\hertz}$. (b) Wigner functions showing the evolution of the cavity state with (bottom) and without (top) the fluxonium qubit, along with fidelity $F$ with the initial state (up to trivial rotations). Without the fluxonium, the self-Kerr induced by the transmon goes through a full revival of the state, via various coherent state superpositions and other nonclassical states. With the fluxonium, the state remains very close to the original coherent state with only small distortions appearing after $\SI{100}{\micro\second}$.}
    \label{fig:coherent}
\end{figure}

The Hamiltonian for the cavity-transmon-fluxonium system is
\begin{equation}
      \begin{split}
          H^{CTF} = &\omega_c a^\dag a + \omega_t b^\dag b  + \frac\chi2 b^\dag b^\dag b b + \lambda_t\left(ab^\dag + a^\dag b \right)
       \\
       &+  \sum_j  \omega_{f,j} \, \ketbra{j_f}{j_f}  + \sum_{j<k} \lambda_{f,jk}\left(\ketbra{k_f}{j_f}a + \ketbra{j_f}{k_f}a^\dag\right),
      \end{split}
\end{equation}
where $\omega_c$ is the cavity frequency, $\omega_t$ is the transmon fundamental frequency, $\chi$ is the transmon nonlinearity, $\omega_{f,j}$ is the energy of the $j$-th level of the fluxonium, $\lambda_{t,jk}$ and $\lambda_{f,jk}$ are the coupling strengths of the $j\to k$ qubit transitions to the cavity mode, $a$ is the cavity annihilation operator, $b$ is the transmon annihilation operator and $\ket{k_{t,f}}$ are the eigenstates of the qubits. The fluxonium energy levels and the relative magnitudes of the coupling constants are determined by the device parameters: the Josephson, charging and inductive energies and the flux through the device.

In Figure \ref{fig:coherent} we show how a coherent state with initial amplitude $\alpha=2$ evolves both with and without the additional fluxonium. We see that when only the transmon is present the cavity state experiences a full revival after approximately \SI{100}{\micro\second}, while adding the fluxonium means that the cavity amplitude is held almost constant for this full period, at least an order of magnitude reduction of the self-Kerr. We also show Wigner functions at various times during the evolution along with their fidelity with the initial condition, which is calculated by
\begin{equation}
      F(t)= \min_{\theta \in {0,2\pi}} \langle \alpha e^{i\theta} \ket{\psi(t)}.
\end{equation}
We see that the self-Kerr in the absence of the fluxonium distorts the state dramatically, producing Schr\"{o}dinger cat states and other coherent state superpositions before returning to the initial coherent state. When the fluxonium is included, the fidelity remains above 0.90 for \SI{50}{\micro\second}, before the state starts to become slightly squeezed. This is likely to be due to the presence of higher-order linearities, which are not perfectly cancelled by the fluxonium. In this case, the ratio of the interaction strengths with and without the fluxonium is approximately 10, rather than the theoretical infinite limit given with two Duffing-like qubits.

\begin{figure}[h!!]
    \centering
    \includegraphics[width=.49\columnwidth,clip=true,trim=.7cm 0cm 2cm 1.5cm]{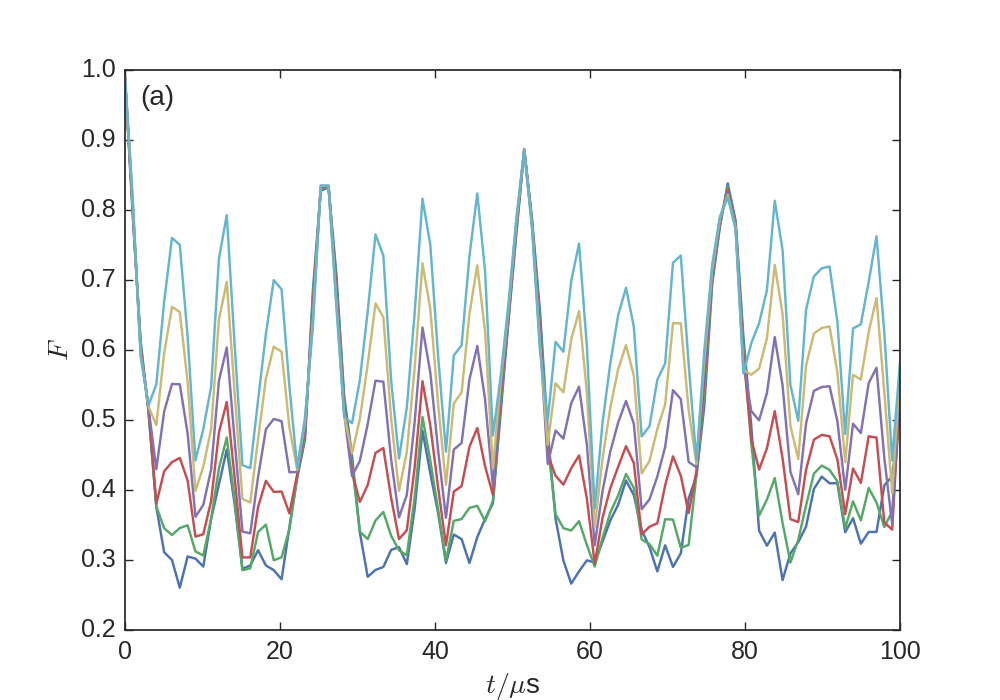}
    \includegraphics[width=.49\columnwidth,clip=true,trim=.7cm 0cm 2cm 1.5cm]{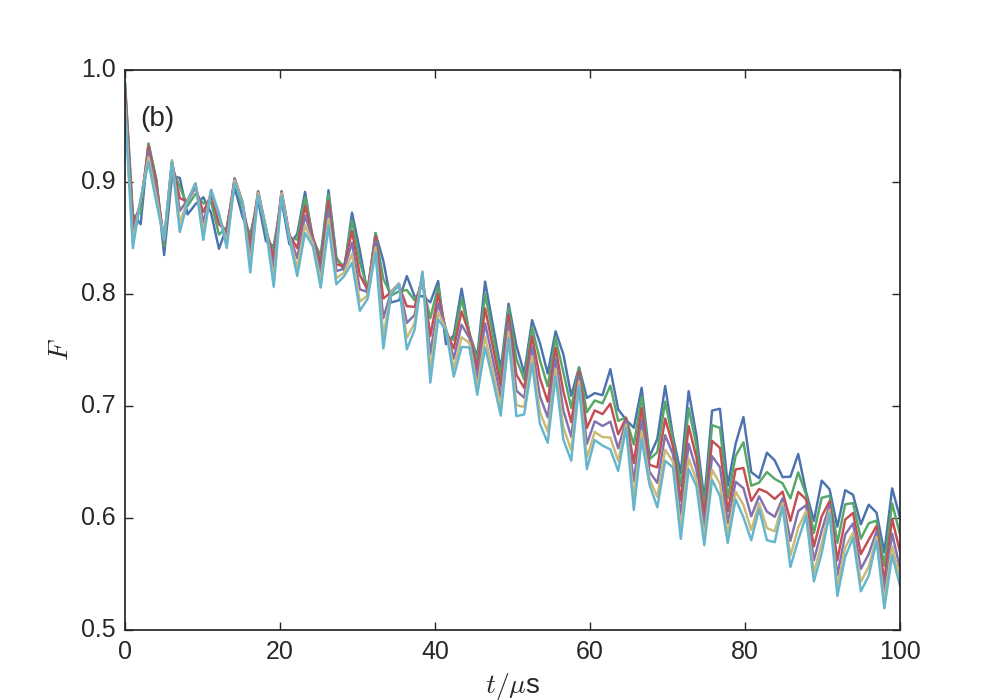}
    \caption{Fidelity plots showing the evolution of a variety of states of the form $|\phi\rangle$ (see text), showing the fidelity with the initial state as a function of time (a) with only a transmon qubit and (b) when coupled to both a transmon and a fluxonium. All the superpositions are preserved with high fidelity for the full duration of 100 $\mu$s, suggesting that all possible logical states can be stored effectively in a quantum memory. The system parameters are $\omega_c=\SI{9.2}{\giga\hertz}$, $\omega_t=\SI{8.2}{\giga\hertz}$, $ \chi=\SI{-300}{\mega \hertz}$, $g_t=\SI{80}{\mega\hertz}$, $\omega_{f,j}=0, 5.505, 10.904, 13.213\si{\giga\hertz}$ and $g_f=\SI{30}{\mega\hertz}$. The initial drop in fidelity from 1 to around 0.9 and subsequent oscillations are caused by the fact that the bare cavity states are not exact eigenstates of the coupled system.}
    \label{fig:fidelity}
\end{figure}

To store quantum information we can encode it in a superposition of coherent states in the cavity field. Instead of the typical Schr\"{o}dinger cat states, we may want to use a space of logical computation states which all have positive parity
\begin{equation}
      \ket{\phi} = a\ket{0^L} + b\ket{1^L} = N_\alpha \Big( a \left(\ket{\alpha}+\ket{-\alpha}\right)+b\left(\ket{i\alpha}+\ket{-i\alpha}\right) \Big),
\end{equation}
where $N^+_\alpha$ is a normalisation factor. These logical qubit states are of interest because the loss of a single photon from the state moves the state into an orthogonal subspace to the computational space. This allows easier reconstruction of the state and makes them more suitable for the storage of quantum information \cite{Ofek2016}. Again, we want to see how these states evolve with and without the fluxonium present. In Fig. \ref{fig:fidelity}, we plot the fidelity as a function of time with initial states with $a={0,0.1,0.2,0.3,0.4,0.5}$, noting that the $a>0.5$ states are just rotations of the $a<0.5$ case. This time fidelities are given by
\begin{equation}
      F(t)= \min_{\theta \in {0,2\pi}} N_\alpha \left(a\left(\bra{e^{i\theta}\alpha}+\bra{-e^{i\theta}\alpha}\right)+b\left(\bra{ie^{i\theta}\alpha}+\bra{-ie^{i\theta}\alpha}\right)\right) \ket{\psi(t)}.
\end{equation}
For these superpositions, collapses and revivals occur much more rapidly than for coherent states. While this means that less time must be waited for a full revival, it also means that this time must be calculated very accurately if it is to be corrected for. Introducing the fluxonium means that all states are preserved with greater than 0.50 fidelity for \SI{100}{\micro\second}, and are preserved with above 90\% fidelity for longer than a full revival with just the transmon present.

These results show that, by using multiple types of superconducting qubit, we can eliminate the distortions caused by self-Kerr \cite{Ofek2016} when using the cavity as a quantum memory. While these are deterministic, it may not be practical to wait for the next full revival before performing an operation on the cavity state, for example to switch the state back into a superconducting qubit by a qc-map operation \cite{Leghtas2013}. There is therefore a benefit associated with reducing the cavity self-Kerr as far as possible compared with the decoherence time by design, while maintaining the number splitting that enables the cavity field to be controlled via a coupled qubit. This scheme can be implemented passively rather than requiring relatively complex controls to implement cancellation, such as using specific phase gates \cite{Heeres2015}.

\section{Cross-Kerr engineering for entangling gates between two cavities}
\label{sec3}
Having demonstrated that the cavity self-Kerr can be modified by the use of multiple qubit types, we turn to performing a similar engineering of the cross-Kerr interaction between two cavities. We start by considering two cavities which are both coupled to a single two-level atom with the Hamiltonian
\begin{equation}
    H^{2C1Q}= \sum_{i=1}^2 \omega_i a_i^\dag a_i +\frac{\omega_q}{2}\sigma_z+\sum_{i=1}^2 g_i(a_i^\dag \sigma_- + a_i\sigma_+),
\end{equation}
where the two cavities have annihilation operators $a_i$, frequencies $\omega_i$ and coupling to the qubit $g_i$. We can also estimate the cross-Kerr effect using the fourth-order perturbation theory. The approximate eigenvalues in the dispersive limit are
\begin{equation}
\begin{split}
    E_{n_1,n_2,g}^{2C1Q}=&\left( \sum_{i=1,2} \left( \omega_i+ \frac{g_i^2}{\Delta_i}+\frac{g_i^2g_i^2}{\Delta_i^2\Delta_{12}}\right) n_i  - \frac{g_i^4}{\Delta_i^3}n_i^2 \right) -2g_1^2g_2^2\frac{\Delta_1+\Delta_2}{\Delta_1^2\Delta_2^2} n_1n_2.
\end{split}
\end{equation}
For the remainder of this section we will be less concerned with the dispersive shifts of the cavity, and will extract the quadratic correction terms
\begin{align}
          S_i^{2C1Q}&=-\frac{g_i^4}{\Delta_i^3}n_i^2,
          \\
          X^{2C1Q}&=-2g_1^2g_2^2\frac{\Delta_1+\Delta_2}{\Delta_1^2\Delta_2^2} n_1n_2,
\end{align}
which describe the self- and cross-Kerr interactions.

As in the case of self-Kerr on a single mode, considering two qubit levels is not sufficient to describe the qualitative behaviour of the system for weakly anharmonic devices. Adding in a third transmon level produces the new expressions
\begin{align}
      S_i^{2C1T}&=  \frac{\chi g_i^4}{\Delta_i^3(2\Delta_i+\chi)},
      \\
      X^{2C1T}&=2g_1^2g_2^2\frac{\chi(\Delta_1+\Delta_2)}{\Delta_1^2\Delta_2^2(\Delta_1+\Delta_2+\chi)},
\end{align}
where again $\chi$ is the transmon anharmonicity. We see that, as with the self-Kerr the sign of the coupling between the two cavity fields is now determined by the sign of the nonlinearity if the qubit is either above or below both cavities.

\subsection{Modifying the cross-Kerr interaction with two Duffing-like qubits}
\label{sec3-1}
We now consider how we can modify the cross-Kerr between the modes by adding a second device, also coupled to both cavities. Specifically, we are interested in whether we can completely switch off the cross-Kerr in some range of parameters. While a single flux-tunable qubit offers some degree of tunability of the frequency, and therefore the cross-Kerr, the range over which the device can be tuned is relatively small. In this configuration, switching off the interaction entirely is difficult as the interaction reduces proportional to $\Delta^{-3}$. In an extended system where we want to perform many such operations in turn, even a factor of ten reduction in the cross-Kerr will produce a significant loss of entanglement between the two cavities in the time it takes to perform a small number of gates elsewhere in the system. A way to engineer greater tunability o the interaction is therefore desirable. The Hamiltonian for the two-cavity-two-qubit system is
\begin{equation}
      H^{2C2D} = \sum_{i=1}^2 \omega_{c,i} a_i^\dag a_i + \sum_{j=1}^2\omega_{q,j} \left(b_j^\dag b_j + \chi_j b_j^\dag b_j^\dag b_j b_j \right)+ \sum_{i,j=1}^2 g_{ij}(a_i b_j^\dag+a_i ^\dag b_j).
\end{equation}
Similar to the self-Kerr calculations in the previous section, the self- and cross-Kerr terms due to both qubits behave additively in the perturbation expansion. The coefficients for this system are
\begin{align}
S_i^{2C2D} & = \frac{\chi_1 g_{i1}^4}{\Delta_{i1}^3(2\Delta_{i1}+\chi_1)}+\frac{\chi_2 g_{i2}^4}{\Delta_{i2}^3(2\Delta_{i2}+\chi_2)},
\\
X^{2C2D} & = 2g_{11}^2g_{21}^2\frac{\chi_1(\Delta_{11}+\Delta_{21})}{\Delta_{11}^2\Delta_{21}^2(\Delta_{11}+\Delta_{21}+\chi_1)}+2g_{12}^2g_{22}^2\frac{\chi_2(\Delta_{12}+\Delta_{22})}{\Delta_{12}^2\Delta_{22}^2(\Delta_{12}+\Delta_{22}+\chi_2)},
\end{align}
where we have defined detunings $\Delta_{ij}=\omega_{c,i}-\omega_{q,j}$ and couplings $g_{ij}$ between the $i$-th cavity and the $j$-th qubit. We can now see that, similar to the self-Kerr, we can choose the system parameters so that there is no cross-Kerr, and without making any of the energy levels of the system degenerate. By simplifying the parameters such as $g_{11}=g_{22}, ~ g_{12}=g_{21}, ~\chi_1=\chi=-\chi_2, ~\Delta_{11}=-\Delta_{22}, ~\Delta_{12}=-\Delta_{21} $, the self- and cross-Kerr terms are equivalent to
\begin{equation}
S_i^{2C2D} = \chi\frac{g_{i1}^4\Delta_{i2}^3(2\Delta_{i2}-\chi)-g_{i2}^4\Delta_{i1}^3(2\Delta_{i1}+\chi)}{\Delta_{i1}^3\Delta_{i2}^3(2\Delta_{i1}+\chi)(2\Delta_{i2}-\chi)}~, ~~~~X^{2C2T}=0.
\end{equation}

\begin{figure}[h!!]
    \centering
    \includegraphics[width=0.6\columnwidth]{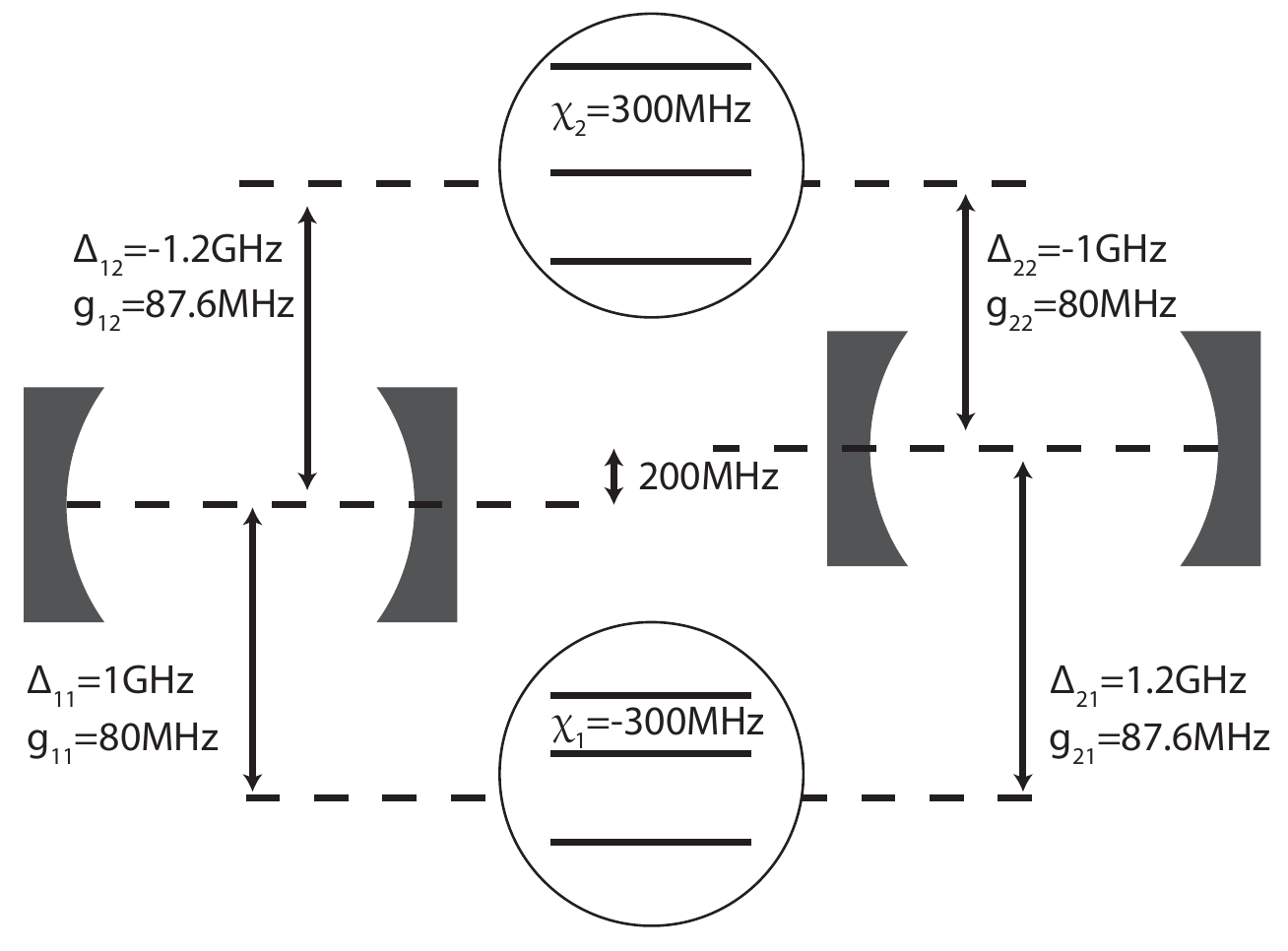}
    \caption{Schematic of setup of two cavities and two Duffing-like qubits that achieves cancellation of the cross-Kerr effect. In this configuration, the degree of entanglement between the two cavity states is unchanged. The entanglement operation can be switched on by detuning the lower qubit further away from the cavities. A change of \SI{1}{\giga\hertz} is sufficient to increase the cross-Kerr almost to the strength it would be if the lower qubit was not present.}
    \label{fig:kerrgate}
\end{figure}

\begin{figure}[h!!]
    \centering
    \includegraphics[width=0.49\columnwidth,clip=]{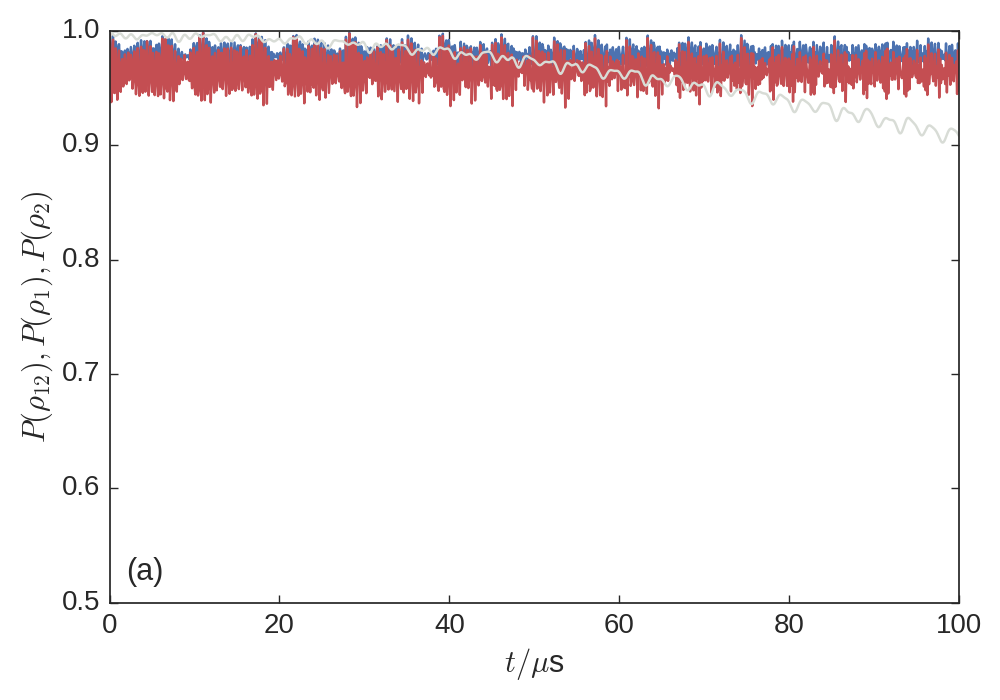}\includegraphics[width=0.49\columnwidth]{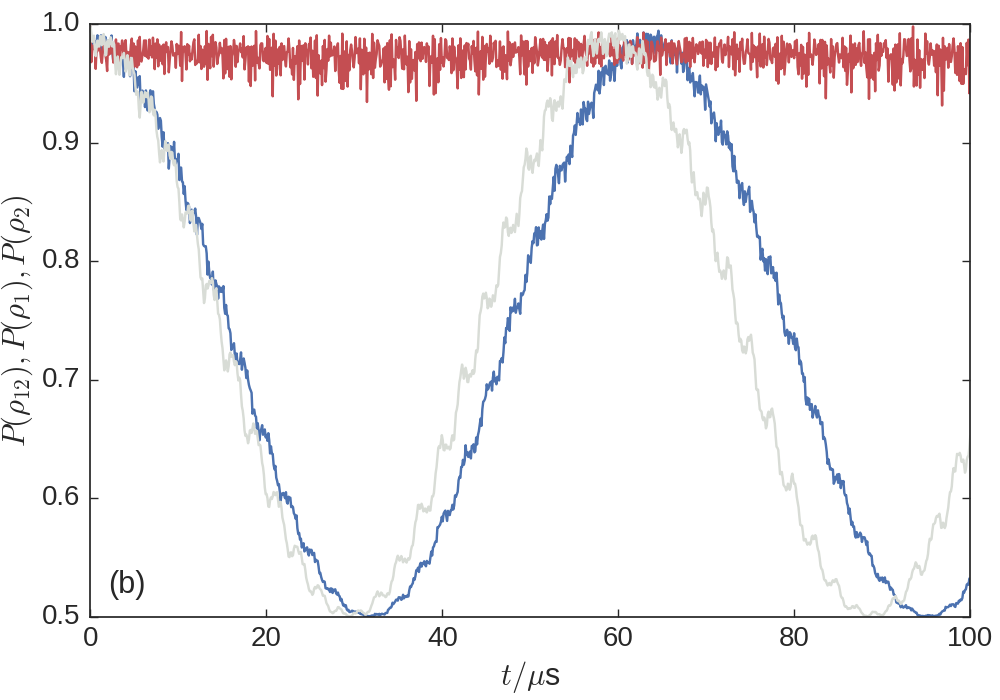}
    \caption{Plots of purities $P(\rho_{12})$ in red and $P(\rho_1)$ in blue showing the effect of switching the cross-Kerr between mode off and on. $P(\rho_2)$ is identical to $P(\rho_1)$. (a) If we tune the second qubit according to the values predicted to the perturbation theory calculations $\Delta_{12}=\SI{-1.2}{\giga\hertz}, \Delta_{21}=\SI{-1}{\giga\hertz}$, then the cavities remain in a product state, shown by both $P(\rho_{12})$ and $P(\rho_1)$ remaining close to 1. Using a single qubit which is detuned by a further \SI{1}{\giga\hertz} from both cavities (shown in grey) does not turn off the entangling operation as effectively. (b) When the second qubit is detuned so that $\Delta_{12}=\SI{-2.2}{\giga\hertz}, \Delta_{21}=\SI{-2}{\giga\hertz}$ then maximal entanglement is achieved after \SI{30}{\micro\second}, almost as quickly as would be achieved if the second device was not present. Other system parameters are $\Delta_{11}=\SI{1}{\giga\hertz}$, $\Delta_{21}=\SI{1200}{\giga\hertz}$, $g_{11}=\SI{80}{\mega\hertz}$, $g_{12}=\SI{87.6}{\mega\hertz}$, $g_{21}=\SI{87.6}{\mega\hertz}$, $g_{22}=\SI{80}{\mega\hertz}$, $\chi_1=\SI{-300}{\mega\hertz}$, $\chi_2=\SI{300}{\mega\hertz}$.}
    \label{fig:gate}
\end{figure}

\subsection{Entangling gates with photons using flux-tunable qubits in two cavities}
\label{sec3-2}
The effect of the cross-Kerr interaction is to create entanglement between the two cavities. When self-Kerr effect and other nonlinearites are present, additional phases are introduced which change what gate is applied but do not change the degree of entanglement present in the system. A useful way to see this is by using the purity \cite{Nielsen} of the state and its individual parts. For a maximally entangled state of two qubits, the full two-qubit system will be pure, whereas tracing out either subsystem will produce a single qubit which is in a completely mixed state.

To perform entangling gates, we now imagine two distinct regimes. In the first, the frequency of the second qubit is selected so that the cross-Kerr is cancelled and the entanglement operation is switched off. In the second the qubit is tuned further from the cavities, $X$ becomes non-zero and entanglement occurs. If this interaction can be switched on for the correct duration, then a maximally entangling operation can be performed. Switching between the two states is achieved by changing the qubit frequency, as can be achieved using an external flux in some transmon devices \cite{Mlynek2014}. To test our perturbation theory results, we fully simulate the evolution of the systems under $H^{2C2T}$, starting with both cavities in intial state given by a superposition of a vacuum and a photon state  $(\ket{0}+\ket{1})/\sqrt{2}$. We show the results of these simulations in Figure \ref{fig:gate}, where we plot the evolution of the system in both the on and off configurations. We plot the purity of the two-qubit state with both cavities traced out $P(\rho_{12}) = tr \left( \rho_{12} \right)$ and the purity of the individual qubit states with the remainder of the system traced out $P(\rho_i) = tr_i \left( \rho_{12} \right)$.

When the qubits are arranged to cancel the cross-Kerr, the system remains in a product state over the full time interval, as shown by the fact that both the two-qubit and single-qubit states have a purity very close to one. The rapid oscillations in the purity are caused by the entanglement of the cavities and qubits that occurs even in the dispersive regime. When one qubit is detuned \SI{1}{\giga\hertz} further from the cavities, we see that the purity of the single qubit states oscillate between 1 and 0.5, with a period of \SI{60}{\micro\second}. This means that if the interaction is switched on for \SI{30}{\micro\second} then the two cavity states become maximally entangled. We compare this to the reduction in the cross-Kerr that can be achieved by detuning a single qubit by \SI{1}{\giga\hertz}, giving an on-off ratio for the interaction of 8. We see that the second qubit allows us to switch off the interaction much more effectively, in principle an infinite on-off ratio, without significantly slowing the entanglement in the `on' configuration. In practise, the range over which a qubit can be tuned is limited and so it is difficult to reduce the cross-Kerr using a single qubit further then shown in Figure \ref{fig:gate}. We therefore have significant advantages over using a single tunable qubit to implement gates.

As with the self-Kerr interaction, it is likely that the present of non-RWA terms will ultimately restrict how much the cross-Kerr can be reduced but, significantly, the controllability of the cavity field via a qubit is maintained, as there is always a qubit close enough that we remain in the dispersive regime. To perform specific logic operations on the two cavities, it may also be possible to improve this scheme by adding extra qubits to each cavity which cancel the self-Kerr. This ability to perform a cross-Kerr based gate between two modes becomes even more valuable when extended to an array of cavities, where it may enable quantum computation when combined with a single qubit rotation implemented via one of the qubits.

\section{Kerr engineering in cavity arrays}
\label{sec4}
\begin{figure}
    \centering
    \includegraphics[width=\columnwidth]{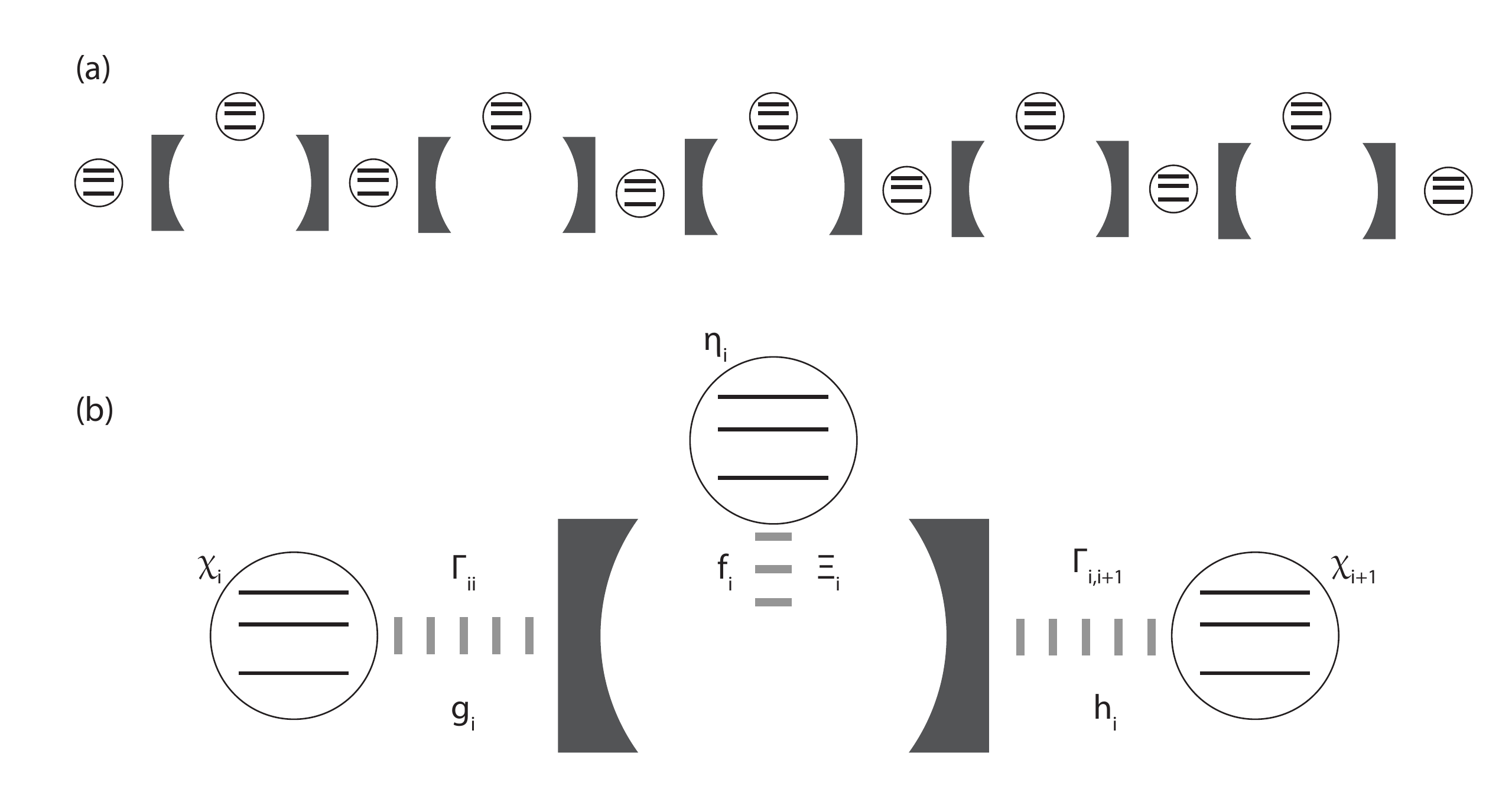}
    \caption{(a) Schematic of a line of superconducting cavities coupled by intermediary qubits and with an on-site qubit per cavity. (b) Enlarged drawing of a single site, showing the relevant couplings and detunings in the model (see main text for definitions of the various parameters).}
    \label{fig:lattice}
\end{figure}

Having demonstrated the principle of Kerr engineering and two applications in smaller systems, we now study a line of $N$ superconducting cavities which are coupled together by intermediary qubits with different nonlinearities. This setup is shown in Figure \ref{fig:lattice}. Each cavity also possesses an on-site qubit, with every device modelled as a Duffing oscillator. The total Hamiltonian for this extended system is

\begin{equation}
      \begin{split}
        H^{tot} = &\sum_i^N \omega_i a_i^\dag a +\sum_i^{N+1}\left(\Omega_j b_i^\dag b_i + \chi_i b_i^\dag b_i^\dag b_i b_i \right) +\sum_i^{N}\left(\tilde{\Omega}_j c_i^\dag c_i + \eta_i c_i^\dag c_i^\dag c_i c_i \right)
    \\
    &+ \sum_i^N\left[ g_i(a_i^\dag b_i + a_i b_i^\dag) + f_i(a_i^\dag c_i+a_i c_i^\dag)+ h_i(a_i^\dag b_{i+1} + a_ib_{i+1}^\dag) \right],
      \end{split}
\end{equation}
where the $b$ operators correspond to the intermediary qubits, while the $c$ operators are for on-site qubits. Each cavity is coupled to three qubits -- two hopping qubits to the left and right, with couplings $g_i$ and $h_i$, and one in the cavity with coupling $f_i$.  Note that the very end qubits are coupled to a single cavity, so we can set $g_1=h_N=0$ for the case of open boundary condition or leave them finite for realising the case of periodic boundary conditions. We also define $\Delta_{ij}=\omega_i-\omega_j$ the detuning between the $i$th and $j$th cavities, $\Gamma_{ij}=\omega_i-\Omega_j$ the coupling between $i$th cavity and $j$th qubit and $\Xi_i=\omega_i-\tilde{\Omega}_i$ the detuning between cavity $i$ and its on-site qubit.

Once again, we apply time-independent perturbation theory to find the self-Kerr on each site $S_i$ and the cross-Kerr between any two cavities $x_{ij}$. These are also derived fully in \ref{app:selfcross}. The self-Kerr is simply the sum of the individual qubits contributions
\begin{equation}
 S_i= \left(\frac{\chi_i g_i^4}{\Gamma_{ii}^3(\Gamma_{ii}+\chi_i)}+\frac{\eta_i f_i^4}{\Xi_i^3(\Xi_i+\eta_i)} +\frac{\chi_{i+1}h_i^4}{\Gamma_{i,i+1}^3(\Gamma_{i,i+1}+\chi_i)}\right)n_i^2,
\end{equation}
and the cross-Kerr between adjacent cavities in the chain is
 \begin{equation}
           X_{i,i+1}=2h_i^2 g_{i+1}^2\frac{\chi_{i+1}(\Gamma_{i,i+1}+\Gamma_{i+1,i+1})}{\Gamma_{i,i+1}^2\Gamma_{i+1,i+1}^2(\Gamma_{i,i+1}+\Gamma_{i+1,i+1}+\chi_{i+1})} n_i n_{i+1},
\end{equation}
while non-adjacent cavities experience no direct interaction up to fourth order in the coupling. We can see from these expressions that by combining different types of qubits (different signs of $\chi_i,\eta_i$) it is possible to arrange the parameters of the chain such that many combinations of self- and cross-Kerr coefficients can be realised and in particular explore the limit of no self-Kerr $S_i=0, X_{i,i+1}\neq 0$. Importantly, it is also possible to choose them such that the dispersive approximation and RWA are obeyed. The existence of the on-site qubits, whose $f_i$ parameter does not feature in the expressions for the cross-Kerr, allows the cross-Kerr to be tuned initially by the intermediary qubits and then the $f_i$ used to achieve the desire self-Kerr. This allows us to realise an effective Hamiltonian
\begin{equation}
    H^{eff} = \sum_i^N \left(\omega'_i a_i^\dag a_i + S_i a_i^\dag a_i^\dag a_i a_i\right) + \sum_i^{N-1} X_{i,i+1} a_i^\dag a_i a_{i+1}^\dag a_{i+1},
\end{equation}
where $\omega'_i$ are rescaled cavity frequencies, $S_i$ are the on-site self-Kerr strengths and $X_{i,i+1}$ is the cross-Kerr between the $i$-th and $i+1$-th cavities. In practice these results only converge if no two cavities or qubits have the same frequency, so selecting the parameters so that the entire system behaves dispersively is non-trivial. However if we inspect two cavities $i,i+n$ then as $n$ increases their respective resonance frequencies can become closer without affecting the fourth order (Kerr) interaction as they only interact in higher orders. In most realistic designs there would be a significant and random fabrication uncertainties but with the advent of tunable qubits and tunable cavities, it is possible to compensate small discrepancies in the nonlinearities by in-situ controlling $\Gamma$ and $\Xi$. This ability to sweep the parameters means that cancellation may be possible even though measuring the bare parameters in an extended is extremely challenging. In addition, parameters of the system may drift or fluctuate in time, necessitating detailed experimental investigation. In principle, the potential of combining different types of qubits can also be extended 2D lattices, although this presents even greater complexity.

\section{Conclusion}
\label{sec5}
We have shown the benefit of combining different types of superconducting qubit to engineer Kerr nonlinearities in systems of cavities. Using perturbation theory and numerical simulations of the exact model we have shown that coupling two types of qubit, with opposite anharmonicities, to a single cavity can enable a complete cancellation of the cavity self-Kerr. When one of the qubits is tunable then the Kerr interaction becomes also tunable. Despite the cancellation of the self-Kerr interaction, the cavity field can still be controlled by the number splitting of the qubit due to the coupling to the cavity. We then showed that a similar result can be achieved for cross-Kerr between modes, demonstrating that configurations exist where the cross-Kerr can be switched on and off, enabling a maximally entangling operation to implemented between the two cavities if frequency-tunable qubits are used. This is a significant improvement over the efficacy of coupling the cavity to a single tunable qubit which be reduces the interaction strength polynomially and therefore cannot remove it entirely without leaving the dispersive regime. Finally, we extended this to a line of superconducting cavities and showed that it could be possible to engineer a chain of cavities with arbitrary on-site and interaction Kerr-coefficients. The on-site qubit for each cavity then allows additional tunability of a class of many-body Hamiltonians and can be used to study phase transitions and quenches out of equilibrium. Implementing cross-Kerr mediated gates in a line of superconducting cavities could also provide a platform for quantum computing.

\section{Acknowledgements}
EG acknowledges funding from EPSRC Grants No. EP/L026082/1 and No. EP/L02263X/1. This work was partially supported by the KIST Institutional Program (Project No. 2E26680-16-P025) and the Overseas Research Program for Young Scientists through Korea Institute for Advanced Study (KIAS). The data underlying this work is available without restriction (doi:10.15126/surreydata.00811086).

\appendix
\section{Perturbation Theory}
\label{app:selfcross}
For system with Hamiltonian
\begin{equation}
    H=H_0+\lambda V,
\end{equation}
where $\lambda$ is a small parameter compared with the terms in the unperturbed Hamiltonian $H_0$, we can write the first four orders of energy corrections in terms of the unperturbed eigenstates $\ket{k}$ and eigenvalues $E^{(0)}_k$ \cite{Meystre}. We define $V_{nm}=\bra{n}V\ket{m}$ and $E_nm=E^{(0)}_n-E^{(0)}_m$. The perturbations are (where all the $k_i$ are summed over with $k_i\neq n$)

\begin{equation}
      \begin{split}
              E^{(1)}_n &= V_{nn}, ~~~   E^{(2)}_n = \frac{\left|V_{ni} \right|^2}{E_{ni}}, ~~~  E^{(3)}_n = \frac{V_{nj}V_{ji}V_{in}}{E_{ni}E_{nj}} -V_{nn}\frac{\left|V_{nj} \right|^2}{E_{nj}^2},
\\
    E^{(4)}_n &= \frac{V_{nk}V_{kj}V_{ji}V_{in}}{E_{ni}E_{nj}E_{nk}} - \frac{\left|V_{nk} \right|^2}{E_{nk}^2}\frac{\left|V_{ni} \right|^2}{E_{ni}} - V_{nn} \frac{V_{nk}V_{kj}V_{jn}}{E_{nj}^2E_{nk}} - V_{nn} \frac{V_{nk}V_{ki}V_{in}}{E_{ni}E_{ni}^2} + V_{nn}^2 \frac{\left|V_{nk} \right|^2}{E_{nk}^3}.
      \end{split}
\end{equation}

In the systems we are looking at, with only linear couplings between cavities and qubits in $V$, applying $V$ to any eigenstate of the unperturbed system will take the system to an orthogonal state, so $V_{nn}=0$. If we are only considering systems without any closed loops, it is also impossible to have terms such as $V_{nk_2}V_{k_2k_1}V_{k_1n}$, with odd numbers of matrix elements. This is because an odd number of applications of the linear coupling term cannot map you back to the original state. This greatly simplifies the perturbed energies to

\begin{equation}
    E^{(1)}_n = 0, ~ ~~   E^{(2)}_n = \frac{\left|V_{ni} \right|^2}{E_{ni}}, ~~~  E^{(3)}_n =0,~~~ E^{(4)}_n = \frac{V_{nk}V_{kj}V_{ji}V_{in}}{E_{ni}E_{nj}E_{nk}} - \frac{\left|V_{nk} \right|^2}{E_{nk}^2}\frac{\left|V_{ni} \right|^2}{E_{ni}}.
\end{equation}

\subsection{Cross-Kerr interaction with two-level qubits}
One case we apply perturbation theory to is two cavities that are both coupled to a single qubit. The Hamiltonian in this case is
\begin{equation}
    H^{2C1Q}= \sum_{i=1}^2 \omega_i a_i^\dag a_i +\frac{\omega_q}{2}\sigma_z+\sum_{i=1}^2 g_i(a_i^\dag \sigma_- + a_i\sigma_+).
\end{equation}
The second order correction to the cavity energy levels includes only terms from paths excitations can take through the system by applying the interaction part of the Hamiltonian twice and that return to the initial configuration. The only such paths are one excitation for either cavity hopping to the qubit and back again, so the correction is given by
\begin{equation}
    E_{n_1,n_2,g}^{2C2Q(2)}=\frac{g_1^2}{\Delta_1}n_1 +\frac{g_2^2}{\Delta_2}n_2.
\end{equation}
The fourth order correction includes all paths made up of four applications of the Hamiltonian, where every intermediary state is distinct from the initial (and final) state. The first two terms come from an excitation hopping from one cavity to the other and back again, with opposite signs because they occur in opposite directions.
\begin{equation}
      \begin{split}
          &E_{n_1,n_2,g}^{2C2Q(4)}=g_1^2g_2^2\frac{n_1(n_2+1)}{\Delta_1^2 \Delta_{12}} -g_1^2g_2^2\frac{n_2(n_1+1)}{\Delta_2^2 \Delta_{12}} - \left(\frac{g_1^2}{\Delta_1}n_1 +\frac{g_2^2}{\Delta_2}n_2\right)\left(\frac{g_1^2}{\Delta_1^2}n_1 +\frac{g_2^2}{\Delta_2^2}n_2\right)
    \\
    &=g_1^2g_2^2n_1n_2\left[\frac{1}{\Delta_1^2 \Delta_{12}} -\frac{1}{\Delta_2^2 \Delta_{12}} -\frac{1}{\Delta_1 \Delta_2^2}-\frac{1}{\Delta_1^2\Delta_2} \right]
    \\
    &+g_1^2g_2^2\left[\frac{n_1}{\Delta_1^2 \Delta_{12}} -\frac{n_2}{\Delta_2^2 \Delta_{12}} \right] - \frac{g_1^4}{\Delta_1^3}n_1^2 - \frac{g_2^4}{\Delta_2^3}n_2^2
    \\
    &=g_1^2g_2^2\frac{\Delta_2^2-\Delta_1^2-\Delta_{12}(\Delta_1+\Delta_2)}{\Delta_1^2\Delta_2^2\Delta_{12}} n_1n_2+g_1^2g_2^2\left[\frac{n_1}{\Delta_1^2 \Delta_{12}} -\frac{n_2}{\Delta_2^2 \Delta_{12}} \right] - \frac{g_1^4}{\Delta_1^3}n_1^2 - \frac{g_2^4}{\Delta_2^3}n_2^2
    \\
    &=-2g_1^2g_2^2\frac{\Delta_1+\Delta_2}{\Delta_1^2\Delta_2^2} n_1n_2+g_1^2g_2^2\left[\frac{n_1}{\Delta_1^2 \Delta_{12}} -\frac{n_2}{\Delta_2^2 \Delta_{12}} \right] - \frac{g_1^4}{\Delta_1^3}n_1^2 - \frac{g_2^4}{\Delta_2^3}n_2^2.
      \end{split}
\end{equation}
We can see both cross-Kerr and self-Kerr terms in the fourth order correction. Without considering additional transmon levels, there exists a configuration where both the self- and cross-Kerr can in principle be eliminated by an appropriate choice of the parameters.

\subsection{Cross-Kerr with transmon levels}

If instead of a qubit we use a transmon modelled as a Duffing oscillator, then the Hamiltonian for a single cavity and qubit is
\begin{equation}
      H^{1C1T} = \omega_c a^\dag a + \omega_q b^\dag b + \chi b^\dag b^\dag b b + g(ab^\dag+a^\dag b).
\end{equation}
There are no additional corrections to the cavity energy levels, with the qubit in the ground state, at second order because it is not possible to access the higher qubit levels and return to the initial state with two applications of the interaction Hamiltonian, so
\begin{equation}
    E_{n,g}^{1C1T(2)}=\frac{g^2 n}{\Delta}.
\end{equation}
At fourth order, however, two excitations can hop into the transmon, giving us the fourth order correction
\begin{equation}
     E_{n,g}^{1C1T(4)}=-\frac{g^4n^2}{\Delta^3} + 2\frac{g^4 n (n-1)}{\Delta^2(2\Delta+\chi)}= -2\frac{g^4}{\Delta^2(2\Delta+\chi)}n + \frac{\chi g^4}{\Delta^3(2\Delta + \chi)}n^2.
\end{equation}

Now in the two cavity-one transmon case the Hamiltonian is
\begin{equation}
      H^{2C1T} = \sum_{i=1}^2 \omega_i a_i^\dag a_i + \omega_q b^\dag b + \chi b^\dag b^\dag b b + \sum_{i=1}^2 g(a_i b^\dag+a_i ^\dag b ).
\end{equation}
Again, the second order correction is unchanged by the extra transmon levels
\begin{equation}
    E_{n_1,n_2,g}^{2C1T(2)}=\frac{g_1^2}{\Delta_1}n_1 +\frac{g_2^2}{\Delta_2}n_2,
\end{equation}
but there are new terms at the fourth order correction. These are associated with two photons from one qubit going into the qubit and back again, and also terms where one from each cavity goes into the qubit and then return to the cavities, in four different orders. The fourth order correction is therefore
\begin{equation}
\begin{split}
    &E_{n_1,n_2,g}^{2C1T(4)}=g_1^2g_2^2\frac{n_1(n_2+1)}{\Delta_1^2 \Delta_{12}} -g_1^2g_2^2\frac{n_2(n_1+1)}{\Delta_2^2 \Delta_{12}} - \left(\frac{g_1^2}{\Delta_1}n_1 +\frac{g_2^2}{\Delta_2}n_2\right)\left(\frac{g_1^2}{\Delta_1^2}n_1 +\frac{g_2^2}{\Delta_2^2}n_2\right)
    \\
    &+ \frac{2g_1^4 n_1(n_1-1)}{\Delta_1^2(2\Delta_1+\chi)} + \frac{2g_2^4 n_2(n_2-1)}{\Delta_2^2(2\Delta_2+\chi)}
    \\
    &+2g_1^2g_2^2\left[\frac{1}{\Delta_1^2(\Delta_1+\Delta_2+\chi)}+ \frac{1}{\Delta_2^2(\Delta_1+\Delta_2+\chi)}+ \frac{2}{\Delta_1\Delta_2(\Delta_1+\Delta_2+\chi))}\right] n_1n_2
    \\
    &=g_1^2g_2^2n_1n_2\left[\frac{1}{\Delta_1^2 \Delta_{12}} -\frac{1}{\Delta_2^2 \Delta_{12}} -\frac{1}{\Delta_1 \Delta_2^2}-\frac{1}{\Delta_1^2\Delta_2} \right]+g_1^2g_2^2\left[\frac{n_1}{\Delta_1^2 \Delta_{12}} -\frac{n_2}{\Delta_2^2 \Delta_{12}} \right]
    \\
    &- \frac{g_1^4}{\Delta_1^3}n_1^2 - \frac{g_2^4}{\Delta_2^3}n_2^2+ \frac{2g_1^4 n_1(n_1-1)}{\Delta_1^2(2\Delta_1+\chi)} + \frac{2g_2^4 n_2(n_2-1)}{\Delta_2^2(2\Delta_2+\chi)}
    +2g_1^2g_2^2\frac{(\Delta_1+\Delta_2)^2}{\Delta_1^2\Delta_2^2(\Delta_1+\Delta_2+\chi)} n_1n_2
    \\
    &=2g_1^2g_2^2\frac{\chi(\Delta_1+\Delta_2)}{\Delta_1^2\Delta_2^2(\Delta_1+\Delta_2+\chi)} n_1n_2+g_1^2g_2^2\left[\frac{n_1}{\Delta_1^2\Delta_{12}} -\frac{n_2}{\Delta_2^2 \Delta_{12}} \right] -2\frac{g_1^2}{\Delta_1^2(\Delta_1+\chi)}n_1
    \\
    &-2\frac{g_2^2}{\Delta_2^2(\Delta_2+\chi)}n_2 +  \frac{\chi g_1^4}{\Delta_1^3(2\Delta_1+\chi)}n_1^2 +  \frac{\chi g_2^4}{\Delta_2^3(2\Delta_2+\chi)}n_2^2.
    \end{split}
\end{equation}

We can repeat this with two Duffing-like devices to gives the self- and cross-Kerr coefficients
\begin{align}
S_i^{2C2D} & = \frac{\chi_1 g_{i1}^4}{\Delta_{i1}^3(2\Delta_{i1}+\chi_1)}+\frac{\chi_2 g_{i2}^4}{\Delta_{i2}^3(2\Delta_{i2}+\chi_2)}
\\
X^{2C2D} & = 2g_{11}^2g_{21}^2\frac{\chi_1(\Delta_{11}+\Delta_{21})}{\Delta_{11}^2\Delta_{21}^2(\Delta_{11}+\Delta_{21}+\chi_1)}+2g_{12}^2g_{22}^2\frac{\chi_2(\Delta_{12}+\Delta_{22})}{\Delta_{12}^2\Delta_{22}^2(\Delta_{12}+\Delta_{22}+\chi_2)}.
\end{align}
When considering for than two qubit levels, it is no longer possible to eliminate self- and cross-Kerr simultaneously.

\subsection{Array of cavities}
Finally, we consider a complete line of cavities, each connected by intermediary qubits, and with each cavity possessing its own on-site qubit. The total Hamiltonian using only two-level qubits is
\begin{multline}
    H^{tot} = \sum_i^N \omega_i a_i^\dag a_i +\sum_i^{N+1}\frac{\Omega_j}{2}\sigma_{z_j}+\sum_i^{N}\frac{\Omega'_j}{2}\tilde{\sigma}_{z_j}
    \\
    + \sum_i^N\left[ g_i(a_i^\dag \sigma_{i-} + a_i\sigma_{i+}) + f_i(a_i^\dag \tilde{\sigma}_{i,i}+a_i\tilde{\sigma}_{i,+})h_i(a_i^\dag \sigma_{i+1,-} + a_i\sigma_{i+1,+}) \right].
\end{multline}

The second order correction is as for the simpler cases above, simply summing the contributions due to each qubit

\begin{equation}
      E_{n_i,g}^{(2)}=\sum_i^N \left(\frac{g_i^2}{\Gamma_{ii}}+\frac{f_i^2}{\Xi_i} +\frac{h_i^2}{\Gamma_{i,i+1}}\right)n_i.
\end{equation}
For the fourth order contribution, there are terms associated with: excitations in each cavity hopping to the cavities to the left and right; two excitations in the same cavity hopping onto different qubits; and two excitations from different points in the line hopping onto different qubits. The full correction is
\begin{equation}
      \begin{split}
          E_{n,g,g}^{(4)}=&\sum_i^N \left(\frac{g_i^2 h_i^2 (\Gamma_{ii}+\Gamma_{i,i+1})}{\Gamma_{ii}^2\Gamma_{i,i+1}^2}
           + \frac{g_i^2 f_i^2 (\Gamma_{ii}+\Xi_i)}{\Gamma_{ii}^2\Xi_i^2}
           + \frac{f_i^2 h_i^2 (\Xi_i+\Gamma_{i,i+1})}{\Xi_i^2\Gamma_{i,i+1}^2}\right)n_i(n_i-1)
           \\
           &+\sum\sum_{i\neq j}\left(\frac{g_i^2 h_j^2 (\Gamma_{ii}+\Gamma_{j,j+1})}{\Gamma_{ii}^2\Gamma_{j,j+1}^2}
           + \frac{g_i^2 f_j^2 (\Gamma_{ii}+\Xi_j)}{\Gamma_{ii}^2\Xi_j^2}
           + \frac{f_i^2 h_j^2 (\Xi_i+\Gamma_{j,j+1})}{\Xi_i^2\Gamma_{j,j+1}^2}\right)n_in_j
           \\
           &+\sum\sum_{i\neq j}\left(\frac{g_i^2 g_j^2 (\Gamma_{ii}+\Gamma_{jj})}{2\Gamma_{ii}^2\Gamma_{jj}^2}
           + \frac{h_i^2 h_j^2 (\Gamma_{i,i+1}+\Gamma_{j,j+1})}{2\Gamma_{i,i+1}^2\Gamma_{j,j+1}}
           + \frac{f_i^2 f_j^2 (\Xi_i+\Xi_j)}{2\Xi_i^2\Xi_j^2}\right)n_in_j
           \\
           &- \sum_{i=1}^{N-1}\left(\frac{h_i^2 g_{i+1}^2 (\Gamma_{i,i+1}+\Gamma_{i+1,i+1})}{\Gamma_{i,i+1}^2\Gamma_{i+1,i+1}^2}\right)n_i n_{i+1}
           \\
         &+\sum_{i=1}^{N-1}h_i^2g_{i+1}^2\left(\frac{n_i(n_{i+1}+1)}{ \Delta_{i,i+1}\Gamma_{i,i+1}^2} -\frac{n_{i+1}(n_i+1)}{\Delta_{i,i+1}\Gamma_{i+1,i+1}^2 }\right)
           \\
          & - \sum_i^N \left(\frac{g_i^2}{\Gamma_{ii}}+\frac{f_i^2}{\Xi_i} +\frac{h_i^2}{\Gamma_{i,i+1}}\right)n_i \sum_j^N \left(\frac{g_j^2}{\Gamma_{jj}^2}+\frac{f_j^2}{\Xi_i^2} +\frac{h_j^2}{\Gamma_{j,j+1}^2}\right)n_j
           \\
           =& -\sum_i^N \left(\frac{g_i^2 h_i^2 (\Gamma_{ii}+\Gamma_{i,i+1})}{\Gamma_{ii}^2\Gamma_{i,i+1}^2}
           + \frac{g_i^2 f_i^2 (\Gamma_{ii}+\Xi_i)}{\Gamma_{ii}^2\Xi_i^2}
           + \frac{f_i^2 h_i^2 (\Xi_i+\Gamma_{i,i+1})}{\Xi_i^2\Gamma_{i,i+1}^2}\right)n_i
          \\
          &- \sum_{i=1}^{N-1}\left(\frac{h_i^2 g_{i+1}^2 (\Gamma_{i,i+1}+\Gamma_{i+1,i+1})}{\Gamma_{i,i+1}^2\Gamma_{i+1,i+1}^2}\right)n_i n_{i+1}
          \\
          &+\sum_{i=1}^{N-1}h_i^2g_{i+1}^2\left(\frac{n_i(n_{i+1}+1)}{ \Delta_{i,i+1}\Gamma_{i,i+1}^2} -\frac{n_{i+1}(n_i+1)}{\Delta_{i,i+1}\Gamma_{i+1,i+1}^2 }\right)
           - \sum_i^N  \left(\frac{g_i^4}{\Gamma_{ii}^3}+\frac{f_i^4}{\Xi_i^3} +\frac{h_i^4}{\Gamma_{i,i+1}^3}\right)n_i^2.
      \end{split}
\end{equation}
As before we can now extract the self-Kerr from the new energies
\begin{equation}
      S_i= -\left(\frac{g_i^4}{\Gamma_{ii}^3}+\frac{f_i^4}{\Xi_i^3} +\frac{h_i^4}{\Gamma_{i,i+1}^3}\right)n_i^2,
\end{equation}
and we see that the total nonlinearity is given by adding the separate induced nonlinearities due to the individual qubits. The cross-Kerr between adjacent cavities is only affected by the qubit coupled between them
 \begin{equation}
           X_{i,i+1}=2h_i^2 g_{i+1}^2\frac{\chi_{i+1}(\Gamma_{i,i+1}+\Gamma_{i+1,i+1})}{\Gamma_{i,i+1}^2\Gamma_{i+1,i+1}^2(\Gamma_{i,i+1}+\Gamma_{i+1,i+1}+\chi_{i+1})} n_i n_{i+1}.
\end{equation}
\subsubsection{Adding the third transmon level}
Finally these calculations can be performed for the full array, considering three levels of the transmon
\begin{equation}
      \begin{split}
        H= &\sum_i^N \omega_i a_i^\dag a_i +\sum_i^{N+1}\left(\Omega_j b_i^\dag b_i + \chi_i b_i^\dag b_i^\dag b_i b_i \right) +\sum_i^{N}\left(\tilde{\Omega}_j c_i^\dag c_i + \eta_i c_i^\dag c_i^\dag c_i c_i \right)
    \\
    &+ \sum_i^N\left[ g_i(a_i^\dag b_i + a_i b_i^\dag) + f_i(a_i^\dag c_i+a_i c_i^\dag)+ h_i(a_i^\dag b_{i+1} + a_ib_{i+1}^\dag) \right].
      \end{split}
\end{equation}
As with the cross-Kerr calculations, the only additional terms come from two excitations, either from the same or adjacent cavities hopping into the same qubit. Including these terms gives the Kerr terms quoted in the main text

\begin{align}
 S_i&= \left(\frac{\chi_i g_i^4}{\Gamma_{ii}^3(\Gamma_{ii}+\chi_i)}+\frac{\eta_i f_i^4}{\Xi_i^3(\Xi_i+\eta_i)} +\frac{\chi_{i+1}h_i^4}{\Gamma_{i,i+1}^3(\Gamma_{i,i+1}+\chi_i)}\right)n_i^2,
\\
           X_{i,i+1}&=2h_i^2 g_{i+1}^2\frac{\chi_{i+1}(\Gamma_{i,i+1}+\Gamma_{i+1,i+1})}{\Gamma_{i,i+1}^2\Gamma_{i+1,i+1}^2(\Gamma_{i,i+1}+\Gamma_{i+1,i+1}+\chi_{i+1})} n_i n_{i+1}.
\end{align}

\subsection{Beyond the rotating wave approximation}
\label{app:RWA}
If we do not apply the rotating wave approximation to the cavity-transmon system then the Hamiltonian is
\begin{equation}
      H^{1C1T} = \omega_c a^\dag a + \omega_q b^\dag b + \chi b^\dag b^\dag b b + g(ab^\dag+a^\dag b + ab + a^\dag b^\dag).
\end{equation}
The effect of the new non-RWA terms is to add a further correction to the energy shifts, which to second order in the interaction are
\begin{equation}
    E_{n,g}^{1C1T(2)}=\frac{g^2 n}{\Delta} + \frac{g^2 (n+1)}{\omega_c + \omega_q}.
\end{equation}
At fourth order, these new interaction terms add a large number of additional perturbation terms
\begin{equation}
\begin{split}
     E_{n,g}^{1C1T(4)}&=-\frac{g^4n^2}{\Delta^3} + 2\frac{g^4 n (n-1)}{\Delta^2(2\Delta+\chi)} - \frac{g^4n(n-1)}{\Delta^2(2\omega_c + \chi)} +  \frac{2g^4n^2}{\Delta^2(2\omega_q+\chi)}
     \\
      &+ \frac{2g^4(n+1)^2}{(2\omega_q+\chi)(\omega_c+\omega_q)^2} + \frac{2g^4(n+1)(n+2)}{(\omega_c+\omega_q)^2(2\omega_c + 2\omega_q + \chi)}
      \\
      & + \frac{g^4(n+1)(n+2)}{2\omega_c(\omega_c+\omega_q)^2} - \frac{g^4(n+1)(n+2)}{2\omega_c \Delta (\omega_c+\omega_q)}
     \end{split}
     \end{equation}
In the regime $\Delta \ll |\omega_c + \omega_q|$, it must also be the case that individually $\omega_c,\omega_q \gg \Delta$. This means that all the new terms that are not seen in the JC model are very small in this limit. For typical experiments, resonators and qubits in the 5-10 \si{\giga\hertz} range are used, meaning that with with detunings of the order of \SI{1}{\giga\hertz} these terms will be at least a factor of 10 smaller. The presence of these terms will also place a theoretical limit on the extent to which self-Kerr interaction can be reduced by adding a second qubit.

\vspace{1cm}

\bibliographystyle{iopart-num}
\bibliography{library}

\providecommand{\newblock}{}
\begin{thebibliography}{10}
\expandafter\ifx\csname url\endcsname\relax
  \def\url#1{{\tt #1}}\fi
\expandafter\ifx\csname urlprefix\endcsname\relax\def\urlprefix{URL }\fi
\providecommand{\eprint}[2][]{\url{#2}}

\bibitem{Friedman2000}
Friedman J~R, Patel V, Chen W, Tolpygo S~K and Lukens J~E 2000 {\em Nature\/}
  {\bf 406} 43

\bibitem{VanderWal2000}
van~der Wal C~H, ter Haar A~C~J, Wilhelm F~K, Schouten R~N, Harmans C~J~P~M,
  Orlando T~P, Lloyd S and Mooij J~E 2000 {\em Science\/} {\bf 290} 773--777

\bibitem{Yan2016}
Yan F, Gustavsson S, Kamal A, Birenbaum J, Sears A~P, Hover D, Gudmundsen T~J,
  Rosenberg D, Samach G, Weber S, Yoder J~L, Orlando T~P, Clarke J, Kerman A~J
  and Oliver W~D 2016 {\em Nat. Commun.\/} {\bf 7} 12964

\bibitem{Steffen2010}
Steffen M, Brito F, Divincenzo D~P, Farinelli M, Keefe G, Ketchen M, Kumar S,
  Milliken F, Rothwell M~B, Rozen J and Koch R~H 2010 {\em J. Phys. Condens.
  Matter\/} {\bf 22}

\bibitem{Martinis2002}
Martinis J~M, Nam S, Aumentado J and Urbina C 2002 {\em Phys. Rev. Lett.\/}
  {\bf 89} 117901

\bibitem{Ansmann2009}
Ansmann M, Wang H, Bialczak R~C, Hofheinz M, Lucero E, Neeley M, Connell A~D~O,
  Sank D, Weides M, Wenner J, Cleland A~N and Martinis J~M 2009 {\em Nature\/}
  {\bf 461} 504--506 ISSN 0028-0836
  \urlprefix\url{http://dx.doi.org/10.1038/nature08363}

\bibitem{Strauch2003}
Strauch F~W, Johnson P~R, Dragt A~J, Lobb C~J, Anderson J~R and Wellstood F~C
  2003 {\em Phys. Rev. Lett.\/} {\bf 91} 167005

\bibitem{Chen2012}
Chen Y, Sank D, Malley P~O, White T, Barends R, Chiaro B, Kelly J, Lucero E,
  Mariantoni M, Megrant A, Neill C, Vainsencher A, Wenner J, Yin Y, Cleland A~N
  and Martinis J~M 2012 {\em Appl. Phys. Lett.\/} {\bf 101} 182601
  (\textit{Preprint} \eprint{arXiv:1209.1781v1})

\bibitem{Nakamura1999}
Nakamura Y, Pashkin Y~A and Tsai J~S 1999 {\em Nature\/} {\bf 398} 786--788

\bibitem{Bouchiat1998}
Bouchiat V, Vion D, Joyez P, Esteve D and Devoret M~H 1998 {\em Phys. Scr.\/}
  {\bf T76} 165

\bibitem{Pashkin2009}
Pashkin Y~A, Yamamoto O~A~T and Tsai Y~N~J~S 2009 {\em Quantum Inf. Process.\/}
  {\bf 8} 55--80

\bibitem{Koch2007}
Koch J, Yu T~M, Gambetta J~M, Houck A~A, Schuster D~I, Majer J, Blais A,
  Devoret M, Girvin S~M and Schoelkopf R~J 2007 {\em Phys. Rev. A\/} {\bf 76}
  042319

\bibitem{Manucharyan2009}
Manucharyan V~E, Koch J, Glazman L~I and Devoret M~H 2009 {\em Science\/} {\bf
  326} 113

\bibitem{Kirchmair2013}
Kirchmair G, Vlastakis B, Leghtas Z, Nigg S~E, Paik H, Ginossar E, Mirrahimi M,
  Frunzio L, Girvin S~M and Schoelkopf R~J 2013 {\em Nature\/} {\bf 495} 205

\bibitem{Leghtas2013}
Leghtas Z, Kirchmair G, Vlastakis B, Devoret M, Schoelkopf R~J and Mirrahimi M
  2013 {\em Phys. Rev. A\/} {\bf 87} 042315

\bibitem{Richer2016}
Richer S and Divincenzo D 2016 {\em Phys. Rev. B\/} {\bf 93} 134501
  (\textit{Preprint} \eprint{arXiv:1511.06138v2})

\bibitem{Kerman2013}
Kerman A~J 2013 {\em New J. Phys.\/} {\bf 15} 123011

\bibitem{Wang2016}
Wang C, Gao Y~Y, Reinhold P, Heeres R~W, Ofek N, Chou K, Axline C, Reagor M,
  Blumoff J, Sliwa K~M, Frunzio L, Girvin S~M, Jiang L, Mirrahimi M, Devoret
  M~H and Schoelkopf R~J 2016 {\em Science\/} {\bf 352} 1087

\bibitem{Vacanti2012}
Vacanti G, Fazio R, Kim M~S, Palma G~M, Paternostro M and Vedral V 2012 {\em
  Phys. Rev. A\/} {\bf 85} 022129

\bibitem{Drummond1980a}
Drummond P~D and Gardiner C~W 1980 {\em J. Phys. A. Math. Gen.\/} {\bf 13} 2353

\bibitem{Drummond1980}
Drummond P~D and Walls D~F 1980 {\em J. Phys. A. Math. Gen.\/} {\bf 13}
  725--741

\bibitem{Boissonneault2009}
Boissonneault M, Gambetta J~M and Blais A 2009 {\em Phys. Rev. A\/} {\bf 79}
  013819

\bibitem{Khan2015}
Khan R, Massel F and Heikkil{\"{a}} T~T 2015 {\em Phys. Rev. A\/} {\bf 91}
  043822

\bibitem{Xiong2016}
Xiong W, Jin D~Y, Qiu Y, Lam C~H and You J~Q 2016 {\em Phys. Rev. A\/} {\bf 93}
  023844

\bibitem{Schuster2007}
Schuster D~I, Houck A~A, Schreier J~A, Wallraff A, Gambetta J~M, Blais A,
  Frunzio L, Majer J, Johnson B~R, Devoret M, Girvin S~M and Schoelkopf R~J
  2007 {\em Nature\/} {\bf 445} 515

\bibitem{Holland2015}
Holland E~T, Vlastakis B, Heeres R~W, Reagor M~J, Vool U, Leghtas Z, Frunzio L,
  Kirchmair G, Devoret M~H, Mirrahimi M and Schoelkopf R~J 2015 {\em Phys. Rev.
  Lett.\/} {\bf 115} 180501

\bibitem{Rebic2009}
Rebi{\'{c}} S, Twamley J and Milburn G~J 2009 {\em Phys. Rev. Lett.\/} {\bf
  103} 150503

\bibitem{Chuang1995}
Chuang I~L and Yamamoto Y 1995 {\em Phys. Rev. A\/} {\bf 52} 3489

\bibitem{Milburn1989}
Milburn G~J 1989 {\em Phys. Rev. Lett.\/} {\bf 62} 2124--2127

\bibitem{Glancy2008}
Glancy S and Vasconcelos H~M 2008 {\em J. Opt. Soc. Am. B\/} {\bf 25} 712

\bibitem{Reagor2016}
Reagor M, Pfaff W, Axline C, Heeres R~W, Ofek N, Sliwa K, Holland E, Wang C,
  Blumoff J, Chou K, Hatridge M~J, Frunzio L, Devoret M~H, Jiang L, Schoelkopf
  R~J, Michael J, Frunzio L, Devoret M~H, Jiang L and Schoelkopf R~J 2016 {\em
  Phys. Rev. B\/} {\bf 94} 014506

\bibitem{Mirrahimi2014}
Mirrahimi M, Leghtas Z, Albert V~V, Touzard S, Schoelkopf R~J, Jiang L and
  Devoret M~H 2014 {\em New J. Phys.\/} {\bf 16} 045014

\bibitem{Vlastakis2013}
Vlastakis B, Kirchmair G, Leghtas Z, Nigg S~E, Frunzio L, Girvin S~M, Mirrahimi
  M, Devoret M and Schoelkopf R~J 2013 {\em Science\/} {\bf 342} 607

\bibitem{Leghtas2015}
Leghtas Z, Touzard S, Pop I~M, Kou A, Vlastakis B, Petrenko A, Sliwa K~M, Narla
  A, Shankar S, Hatridge M~J, Reagor M, Frunzio L, Schoelkopf R~J, Mirrahimi M
  and Devoret M~H 2015 {\em Science\/} {\bf 347} 853

\bibitem{Ofek2016}
Ofek N, Petrenko A, Heeres R, Reinhold P, Leghtas Z, Vlastakis B, Liu Y,
  Frunzio L, Girvin S~M, Jiang L, Mirrahimi M, Devoret M~H and Schoelkopf R~J
  2016 {\em Nature\/} {\bf 536} 441

\bibitem{Pfaff2016}
Pfaff W, Axline C~J, Burkhart L~D, Vool U, Reinhold P, Frunzio L, Jiang L,
  Devoret M~H and Schoelkopf R~J 2017 {\em Nat. Phys.\/} {\bf 10} 4143

\bibitem{Juliusson2016}
Juliusson K, Bernon S, Zhou X, Schmitt V, le~Sueur H, Bertet P, Vion D,
  Mirahimi M, Rouchon P and Esteve D 2016 {\em Phys. Rev. A\/} {\bf 94} 063861

\bibitem{Brod2016}
Brod D~J and Combes J 2016 {\em Phys. Rev. Lett.\/} {\bf 117} 080502

\bibitem{Stassi2015}
Stassi R, Liberato S~D, Garziano L, Spagnolo B and Savasta S 2015 {\em Phys.
  Rev. A\/} {\bf 92} 013830

\bibitem{Georgescu2014}
Georgescu I~M, Ashhab S and Nori F 2014 {\em Rev. Mod. Phys.\/} {\bf 86} 153

\bibitem{Jin2013}
Jin J, Rossini D, Fazio R, Leib M and Hartmann M~J 2013 {\em Phys. Rev.
  Lett.\/} {\bf 110} 163605

\bibitem{Creatore2014}
Creatore C, Fazio R, Keeling J and T{\"{u}}reci H~E 2014 {\em Proc. R. Soc.
  A\/} {\bf 470} 20140328

\bibitem{Houck2012}
Houck A~a, T{\"{u}}reci H~E and Koch J 2012 {\em Nat. Phys.\/} {\bf 8} 292

\bibitem{Gersch1963}
Gersch H~A and Knollman G~C 1963 {\em Phys. Rev.\/} {\bf 129} 959

\bibitem{Freericks1994}
Freericks J~K and Monien H 1994 {\em Europhys. Lett.\/} {\bf 26} 545

\bibitem{Leib2010}
Leib M and Hartmann M~J 2010 {\em New J. Phys.\/} {\bf 12} 093031

\bibitem{Zhu2013}
Zhu G, Schmidt S and Koch J 2013 {\em New J. Phys.\/} {\bf 15} 115002

\bibitem{Nissen2012}
Nissen F, Schmidt S, Biondi M, Blatter G, T{\"{u}}reci H~E and Keeling J 2012
  {\em Phys. Rev. Lett.\/} {\bf 108} 233603

\bibitem{Jin2014}
Jin J, Rossini D, Leib M, Hartmann M~J and Fazio R 2014 {\em Phys. Rev. A\/}
  {\bf 90} 023827

\bibitem{Fitzpatrick2016}
Fitzpatrick M, Sundaresan N~M, Li A~C~Y, Koch J and Houck A~A 2017 {\em Phys.
  Rev. X\/} {\bf 7}(1) 011016

\bibitem{Calabrese2007}
Calabrese P and Cardy J 2007 {\em J. Stat. Mech.\/}  P06008

\bibitem{Zueco2009}
Zueco D, Reuther G~M, Kohler S and H{\"{a}}nggi P 2009 {\em Phys. Rev. A\/}
  {\bf 80} 033846

\bibitem{Rabi1937}
Rabi I~I 1937 {\em Phys. Rev.\/} {\bf 51} 652--654 ISSN 0031899X
  (\textit{Preprint} \eprint{arXiv:1011.1669v3})

\bibitem{Irish2005}
Irish E~K, Martin I and Schwab K~C 2005 {\em Phys. Rev. B\/} {\bf 72} 195410

\bibitem{Carbonaro1979}
Carbonaro P, Compagno G and Persico F 1979 {\em Phys. Lett. A\/} {\bf 73} 97

\bibitem{Meystre}
Meystre P and Sargent M 2007 {\em {Elements of Quantum Optics}\/} (Springer)

\bibitem{Garraway2011}
Garraway B~M 2011 {\em Philos. Trans. R. Soc. A\/} {\bf 369} 1137

\bibitem{Stern2014}
Stern M, Catelani G, Kubo Y, Grezes C, Bienfait A, Vion D, Esteve D and Bertet
  P 2014 {\em Phys. Rev. Lett.\/} {\bf 113} 123601

\bibitem{Murch2012}
Murch K~W, Ginossar E, Weber S~J, Vijay R, Girvin S~M and Siddiqi I 2012 {\em
  Phys. Rev. B\/} {\bf 86} 220503

\bibitem{Joo2015}
Joo J and Ginossar E 2015 {\em Sci. Rep.\/} {\bf 6} 26338

\bibitem{Heeres2015}
Heeres R~W, Vlastakis B, Holland E, Krastanov S, Albert V~V, Frunzio L, Jiang L
  and Schoelkopf R~J 2015 {\em Phys. Rev. Lett.\/} {\bf 115} 137002

\bibitem{Nielsen}
Nielsen M~A and Chuang I~L 2011 {\em {Quantum Computation and Quantum
  Information}\/} 10th ed (Cambridge University Press) ISBN 1107002176

\bibitem{Mlynek2014}
Mlynek J~A, Abdumalikov A~A, Eichler C and Wallraff A 2014 {\em Nat. Commun.\/}
  {\bf 5} 5186

\end{thebibliography}

\end{document}